\documentclass[preprint,a4paper]{revtex4-2}
\setcitestyle{super}

\usepackage[utf8]{inputenc}
\usepackage{amsmath,amsfonts,amssymb}
\usepackage{geometry}
\usepackage{graphicx}
\usepackage{color}
\usepackage{xcolor}
\usepackage{empheq}
\usepackage{hyperref}
\usepackage{bm}
\usepackage{url}
\usepackage{diagbox}
\usepackage{array}
\usepackage{setspace}
\usepackage{multirow}
 \usepackage{amstext,amsmath,graphicx,bm}

\newcommand*{\alex}[1]{\textcolor{black}{#1}}
\newcommand*{\alexx}[1]{\textcolor{black}{#1}}
\newcommand*{\corr}[1]{\textcolor{black}{#1}}
\newcommand*{\rev}[1]{\textcolor{black}{#1}}

\newcommand*{\Paul}[1]{\textcolor{black}{#1}}

\def\r'{\mathbf{r'}}

\def\uin{\mathbf{u}_\textrm{in}} 
\def\rp{\mathbf{r_p}} 

\begin{abstract}

\textbf{Label-free microscopy exploits light scattering to obtain a three-dimensional image of biological tissues. However, light propagation is affected by aberrations and multiple scattering, which drastically degrade the image quality and limit the penetration depth. Multi-conjugate adaptive optics and time-gated matrix approaches have been developed to compensate for aberrations but the associated frame rate is extremely limited for 3D imaging. 
Here we develop a multi-spectral matrix approach to solve these fundamental problems. Based on \corr{a sparse illumination scheme and} an interferometric measurement of the \corr{reflected wave-field at multiple wavelengths}, the focusing process can be optimized in post-processing \corr{for} any voxel by addressing independently each frequency component of the \corr{reflection matrix}. A proof-of-concept experiment demonstrates the three-dimensional image of an opaque human cornea over a 0.1 mm$^3$-field-of-view at a {290} nm-resolution and a 1 Hz-frame rate. This work  paves the way towards a fully-digital microscope allowing real-time, in-vivo, quantitative and deep inspection of tissues.}
\end{abstract}

\begin{document}

\title{Multi-Spectral Reflection Matrix for\\ Ultra-Fast 3D Label-Free Microscopy}

\author         {Paul Balondrade{$^\dag$}}
\author         {Victor Barolle{$^\dag$}}
\author         {Nicolas Guigui}
\author         {Emeric Auriant}
\author         {Nathan Rougier}
\author         {Claude Boccara}
\author         {Mathias Fink}
\author         {Alexandre Aubry{$^*$}}
\affiliation    {Institut Langevin, ESPCI Paris, PSL University, CNRS, 75005 Paris, France\\ \textnormal{$^\dag$ These authors equally contributed to this work.\\
{$^*$}Corresponding author (e-mail: alexandre.aubry@espci.fr)}}

\date{\today}
   \maketitle

   \clearpage 

\noindent {\large \textbf{Introduction}} 

Imaging of thick scattering tissues remains the greatest challenge in label-free microscopy \cite{ntziachristos_going_2010,gigan_optical_2017,Bertolotti2022}. On the one hand, short-scale inhomogeneities of the refractive index backscatter light and the reflected wave-field can be leveraged to provide a structural image of the sample. On the other hand, larger-scale inhomogeneities give rise to forward multiple scattering events that distort the incident and reflected wave-fronts. This phenomenon, known as aberrations, leads to a drastic degradation of resolution and contrast at depths greater than the scattering mean free path $\ell_s$ ($\sim$100 $\mu$m in biological tissues). 

To circumvent this issue, adaptive optics (AO) has been transposed from astronomy to microscopy for the last twenty years~\cite{booth_adaptive_2014}. The basic idea is to compensate for wave distortions either by a direct sampling of the wave-field generated by a guide star or by an indirect metric optimization of the image. Unfortunately, AO correction is limited to a finite area, the so-called isoplanatic patch, the area over which aberrations can be considered spatially invariant. This problem becomes particularly important for deep imaging, where each isoplanatic patch reduces to a speckle grain at depths larger than the transport mean free path $\ell_t$ ($\sim 1$ mm in biological tissues). Multi-conjugate AO could increase the corrected field-of-view~\cite{Wu2015}, but this would be at the price of a much more complex optical setup and an extremely long optimization process~\cite{Park2018}.

More recently, following seminal works that proposed post-processing computational strategies for AO~\cite{Ralston2007,Adie2012,Ahmad2013,Hillmann2016,Katz2023}, a reflection matrix approach has been developed for deep imaging~\cite{kang_imaging_2015,badon_smart_2016,kang_high-resolution_2017,badon_distortion_2020,yoon_laser_2020,Kwon2023,Najar2023}. The basic idea is to illuminate the sample by a set of input wave-fronts and record via interferometry the reflected wave-front on a camera. Once this reflection matrix is measured, a set of matrix operations can be applied in order to perform a local compensation of aberrations and restore a diffraction-limited resolution for each pixel of the field-of-view. Nevertheless, the existing approaches suffer from several limitations. In most experimental works~\cite{kang_imaging_2015,badon_smart_2016,badon_distortion_2020,yoon_laser_2020,Kwon2023,Najar2023}, the reflection matrix is time-gated around the ballistic time as usually performed in time-domain OCT~\cite{Huang_91}. Such a measurement has one main advantage since it enables the temporal filtering of most multiply-scattered photons~\cite{badon_multiple_2017}. However, it also suffers from two strong drawbacks. First, time-gating means that a large part of the information on the medium is discarded: Only the weakly distorted paths are recorded and can be compensated by a spatial phase modulation of the incident and reflected wave-fronts. 
Second, volumetric imaging can only be obtained by a mechanical axial scanning of the sample, which limits the frame rate {$Fps$} to, at best, $10^6$ voxels.$s^{-1}$ for a high quality correction over millimetric FOVs. 

To go beyond, an acquisition of a spectral reflection matrix is required in order to capture all the information required for the \alex{three-dimensional} imaging of a sample. \alex{In recent works~\cite{Lee2023,Zhang2023}, the spatio-temporal degrees of freedom exhibited by the reflection matrix have been exploited for tailoring dispersive focusing laws. However, the acquisition rate was slow ({$Fps$} $\sim 10^3$ voxels.s$^{-1}$) \alex{because the number of input wave-fronts scaled as the number of voxels in the image}. Moreover, the experimental demonstration was limited to the imaging of a resolution target through a scattering medium~\cite{Lee2023,Zhang2023} or a sparse medium made of colloidal particles~\cite{Zhang2023}. In this paper, we go beyond an academic proof-of-concept and address the extremely challenging case of ultra-fast 3D imaging of biological tissues themselves (nerves, cells, collagen, extracellular matrix etc.). \alex{In particular, we will show how the number of input wave-fronts can be drastically decreased by deterministic focusing operations applied to the reflection matrix guided by a self-portrait of the focusing process.}}

\alex{To that aim, }we report on a measurement of the multi-spectral reflection matrix at a much higher frame rate (${Fps} \sim \corr{5} .10^{9}$ voxels.s$^{-1}$), with a 3D imaging demonstration on an \textit{ex-vivo} opaque cornea at a resolution of \alex{0.29} $\mu$m and \alex{0.5} $\mu$m in the transverse and axial directions, respectively.
The experimental set up combines a \alex{Fourier-domain full-field OCT (FD-FF-OCT)} setup~\cite{povazay_full-field_2006, hillmann_efficient_2012,Auksorius2020} with a coherent multi-illumination scheme. Capable of recording a polychromatic reflection matrix of {$10^{10}$} coefficients in less than 1 s with an ultra-fast camera, this device is fully compatible with future \textit{in-vivo} applications. As in \alex{FD-FF-OCT}, a \rev{frequency} Fourier transform and numerical refocusing can provide a 3D image of the sample for each incident wave-front~\cite{povazay_full-field_2006, hillmann_efficient_2012,Auksorius2020} but, as expected, multiple scattering is shown to strongly hamper the imaging process. A coherent compound of images obtained for each illumination in post-processing can then provide a digital confocal image but its resolution and contrast are drastically affected by sample-induced aberrations. Interestingly, reflection matrix imaging (RMI) can go beyond by decoupling input and output focusing points at each time-of-flight. \corr{A focused \rev{and time-gated} reflection matrix is synthesized\rev{. It contains the} impulse responses between virtual sources and detectors conjugated with each voxel inside the sample.}
While previous works only considered \corr{focusing points} at the same depth~\cite{badon_smart_2016,kang_high-resolution_2017,badon_distortion_2020,yoon_laser_2020}, we show here that their axial scan gives access to a self-portrait of light propagation inside the sample. \corr{The focusing process can then be guided by this reflection point spread function, thereby allowing an optimized compensation of \rev{defocus} aberrations} at each depth of the sample\rev{, even in presence of multiple scattering}. \corr{Finally, transverse aberrations can be tackled by means of a local analysis of wave distortions, as initially proposed with ultrasound~\cite{Bureau2023} and recently transposed to optics~\cite{Najar2023}. However, while those previous works implied a complex multi-scale analysis of wave distortions, the prior compensation of \rev{defocus} aberrations here enables a direct and local correction of transverse aberrations.} A digital clearing of refractive index heterogeneities is thus applied and a three-dimensional image of the sample is obtained with an optimized contrast and close-to-ideal resolution throughout the volume. \rev{The ability of RMI in addressing forward multiple scattering paths is finally investigated by considering the imaging of a resolution target through a more opaque region of the cornea.} 

\clearpage

\noindent {\large \textbf{Results}}

\noindent {\textbf{Recording the Multi-Spectral Reflection Matrix.}} 

\begin{figure}
    \includegraphics[width=\textwidth]{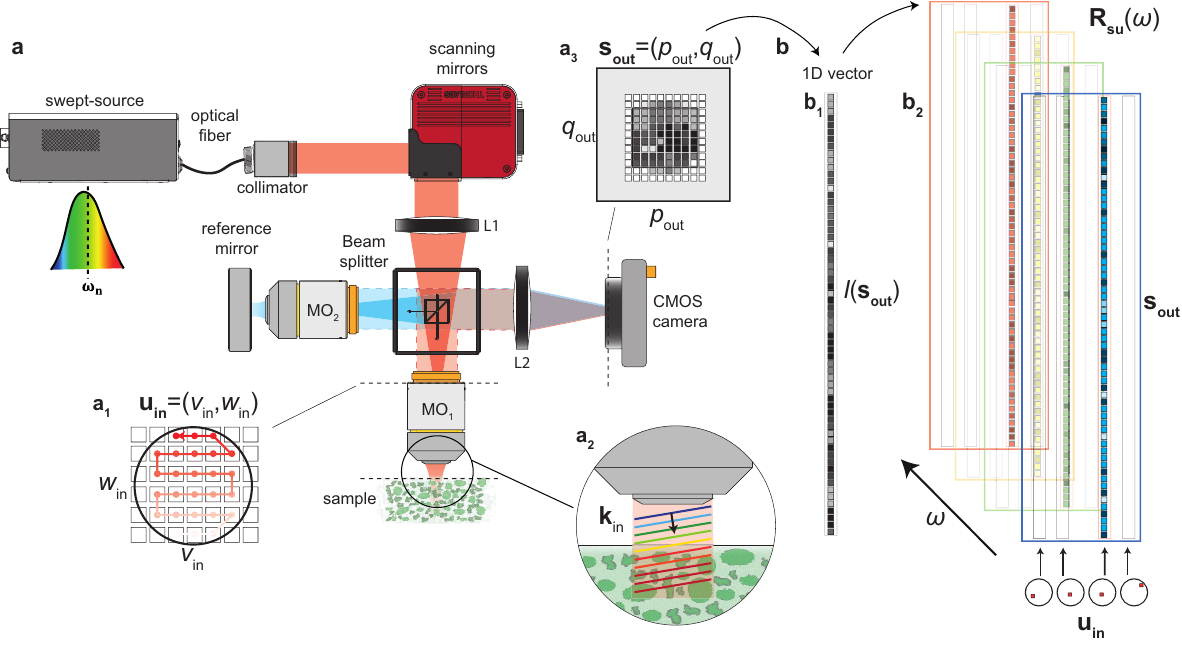}
    \caption{\textbf{Measuring the multi-spectral reflection matrix.} \textbf{a}, A \alex{wavelength swept light source} illuminates a Linnik interferometer through a collimator, two scanning mirrors and a lens (L1) that allows a raster scanning of the focal spot in the MO pupil planes ($\mathbf{u}_\textrm{in}$) in each arm (\textbf{a$_1$}). The sample \alex{placed} in the focal plane of the first MO ({MO$_1$}, NA=0.8) is thus illuminated by a set of plane waves at each frequency of the light source bandwidth ({\textbf{a$_2$}}). The backscattered wave field is collected through the same MO, focused by means of a second lens L2 on the surface of a CMOS camera where it interferes with a reference beam ({\textbf{a$_3$}}). The latter beam results from the reflection of the same incident wave-fronts by a reference mirror placed in the focal plane of the second MO ({MO$_2$}, NA=0.8). \textbf{b}, At each frequency $\omega$, for each input wave-front $\mathbf{u}_\textrm{in}$, the interferogram $I(\mathbf{s}_\textrm{out})$ (\textbf{b$_1$}) recorded by each pixel $\mathbf{s}_\textrm{out}$ of the camera provides one column of the spectral reflection matrix $\mathbf{R}_{\mathbf{su}}(\omega)=[R(\mathbf{s}_{\textrm{out}},\mathbf{u}_{\textrm{in}},\omega)]$ (\textbf{b$_2$}).}
    \label{fig:setup}
\end{figure}

3D matrix imaging is based on the measurement of a multi-spectral reflection matrix from the scattering sample. The experimental setup and procedure are described in Fig.~\ref{fig:setup} (see Methods and Supplementary Figure S1). Inspired by \rev{FD-FF-OCT}~\cite{stremplewski_vivo_2019}, it simply consists in a Linnik interferometer (Fig.~\ref{fig:setup}a). In the first arm, a reference mirror is placed in the focal plane of a microscope objective (MO). The second arm contains the scattering sample to be imaged through an identical MO. This interferometer is illuminated by a swept source through two scanning mirrors and a lens that allows a raster scanning of the \rev{focused beam} in the MO pupil planes (Fig.~\ref{fig:setup}a$_1$). The sample and reference mirror are thus illuminated by a set of plane waves at each frequency of the light source bandwidth  (Fig.~\ref{fig:setup}a$_2$). The reflected waves are collected through the same MOs and, ultimately, interfere on a camera conjugated with the focal plane. \corr{For each input wave-front of frequency $\omega$ and coordinate $\mathbf{u}_\textrm{in}$ in the pupil plane, an interferogram $I(\mathbf{s}_\textrm{out},\mathbf{u}_\textrm{in},\omega)$ is recorded by the camera whose sensors are identified by their transverse position $\mathbf{s}_\textrm{out}$ (Fig.~\ref{fig:setup}a$_3$). Each interferogram provides one column of the multi-spectral reflection matrix $\mathbf{R}_{\mathbf{su}}(\omega)=\left [R(\mathbf{s}_\textrm{out},\mathbf{u}_\textrm{in},\omega) \right ]$ (see Methods and Fig.~\ref{fig:setup}b). The subscripts associated with $\mathbf{R}$ here denote the output and input basis in which the reflection matrix is defined. Indeed, this matrix will be projected between different bases in this work: (\textit{i}) the camera sensor basis ($\mathbf{s}$); (\textit{ii}) the plane wave basis ($\mathbf{u}$) and; (\textit{iii}) the focused basis ($\bm{\rho}$), in which the image is built. The coefficients of the multi-spectral reflection matrix, $R(\mathbf{s}_\textrm{out},\mathbf{u}_\textrm{in},\omega)$, are noted as functions of the coordinates in the planes associated with each basis. The same convention will be considered for the other operators considered in this paper.} 

In the opaque cornea experiment, the reflection matrix $\mathbf{R}$ is measured with $N_{\textrm{in}} = 177$ plane waves, corresponding to a full scan of the immersion MO pupil (NA=0.8, refractive index $n_0=1.33$). The interferograms are recorded by $N_{\textrm{out}}=1024^2$ pixels of the camera, corresponding to an output FOV of $\Omega_{\textrm{out}} \times \Omega_{\textrm{out}} = 297 \times 297$ $\mu$m$^2$, with a spatial sampling $\delta \rho_{\textrm{out}} = 290$ nm. Finally, $N_{\omega}=201$ independent frequencies are used to probe the sample within the frequency bandwidth $[800\corr{,}875]$ nm of the light source. All the information about the sample is thus contained in the \corr{$2.10^{9}$} coefficients acquired in 1.4 s. In the following, we show how to post-process this wealth of optical data to build a 3D highly-contrasted image of the cornea at a diffraction-limited resolution. 

\vspace{5 mm}

\noindent {\textbf{Ultra-fast Three-Dimensional Imaging.}} 

To that aim, the most direct path is to perform, a Fourier transform  over frequency $\omega$ of the back-scattered wave-field recorded for one \corr{plane-wave} illumination~\cite{povazay_full-field_2006}: This is the principle of \rev{full-field swept-source} OCT which provides an image \alex{whose axial dimension is dictated by photons' times-of-flight (Supplementary Figure S2).} \alex{The resulting image is, however, completely blurred without any connection with the sample reflectivity. Indeed, a high NA implies} a very restricted depth-of-field which is prohibitory for 3D imaging~\cite{Sentenac2018}: \corr{$\delta z_f~\sim~ 2 n_0 \lambda /NA^2 \sim {3.5}$ $\mu$m, with $\lambda$ the central wavelength}. A prior numerical focusing of the wave-field recorded by the camera shall be performed at each depth $z$ of the sample. This is the principle of the holoscope developed by Hillmann \textit{et al.} about a decade ago~\cite{Hillmann2016}.

\begin{figure}
    \includegraphics[width=\textwidth]{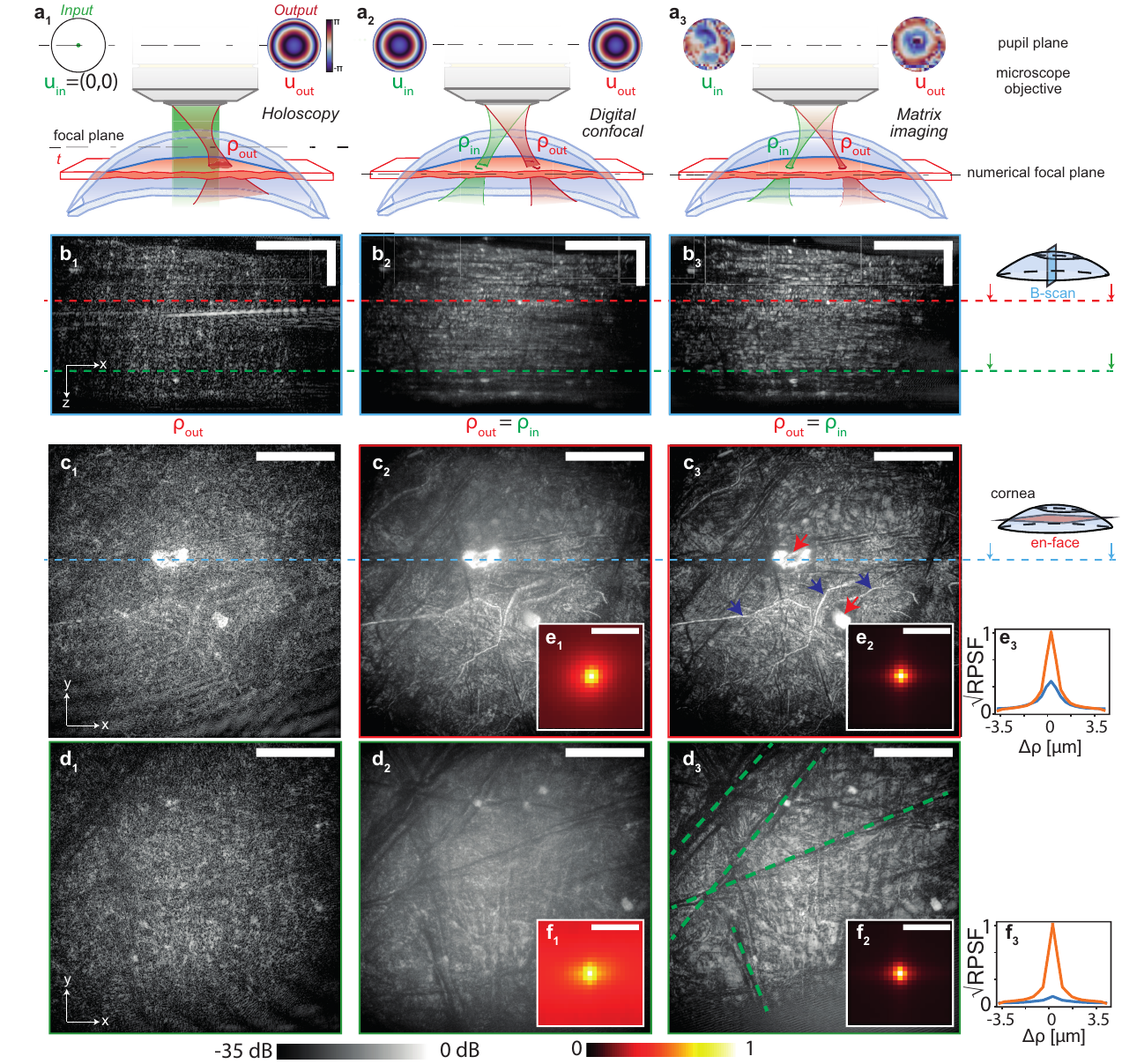}
    \caption{\textbf{From Holoscopy to Matrix Imaging.} \textbf{a}, Imaging Methods. \textbf{a$_1$}, Holoscopy:  The sample is illuminated by a plane wave (in green) and an image is produced by spatio-temporal focusing of the back-scattered wave-field on each voxel \corr{$(\bm{\rho}_\textrm{out},z)$} mapping the sample (red). \textbf{a$_2$}, Digital confocal microscopy (DCM): The sample is illuminated by a set of plane waves (in green) and a focused refection matrix $\mathbf{R}_{\bm{\rho}\bm{\rho}}(z_t)$ is built by numerical focusing. A 3D confocal image is deduced from the diagonal elements of $\mathbf{R}_{\bm{\rho}\bm{\rho}}(\corr{z_t})$ at each depth $z_t$. \textbf{a$_3$}, Reflection matrix imaging (RMI): A local compensation of wave distortions is performed for each voxel. \textbf{b}, B-scan image showing one longitudinal section of the cornea reflectivity.  \textbf{c}-\textbf{d}, En-face image of the cornea at \corr{$z_t=105$ $\mu$m and 230 $\mu$m}, respectively [scale bar: 75 $\mu$m]. In panels (\textbf{a})-(\textbf{d}), subscripts 1, 2 and 3 stand for holoscopy, DCM and RMI, respectively. \alexx{In panel c$_3$, blue and red arrows design some corneal nerves and keratocytes , respectively; in panel d$_3$, stroma striae are highlighted by green dashed lines.}  \textbf{e}, RPSF  at depth \corr{$z_t=105$} $\mu$m [scale bar: 3 $\mu$m] for DCM (\textbf{e$_1$}-\textbf{f$_1$}) and RMI (\textbf{e$_2$}-\textbf{f$_2$}) images. \corr{The color bar is in a square root scale.} The radial evolutions of these RPSFs are compared in panel (\textbf{e$_3$}) [DCM: blue; RMI: orange]. \textbf{f}, Same as in panel (\textbf{e}) but at depth \corr{$z_t=230$} $\mu$m.}
    \label{fig:image}
\end{figure}

This numerical focusing process is performed by means of Fresnel propagators. For this purpose, the multi-spectral reflection matrix \alexx{is projected at output in the focused basis:
\begin{equation}
\label{dualR}
\mathbf{R}_{\bm{\rho} \mathbf{u}}(z,\omega)=\mathbf{F}_{\corr{\bm{\rho}}\mathbf{s}}^{\alex{*}}(z,\omega) \times \mathbf{R}_{\mathbf{su}}(\omega)
\end{equation}
where the symbol\corr{s} $*$ \corr{and $\times$ stand for the phase conjugation operation and the standard matrix product, respectively}. $\mathbf{F}_{\corr{\bm{\rho}}\mathbf{s}}(z,\omega)=[F(\corr{\bm{\rho}_\textrm{out}},\mathbf{s}_\textrm{out},z,\omega)]$ is the Fresnel operator that describes free-space propagation from the camera ($\mathbf{s}$) to any focal plane ($\bm{\rho}$) located at expected depth $z$ in the sample (Methods, \rev{Eq.~\ref{eq22}}).}
\alex{Each frequency component of $\mathbf{R}_{\bm{\rho}\mathbf{u}}(z,\omega)$ should then be recombined in order to time gate the singly-scattered photons. In practice,} an inverse Fourier transform over frequency $\omega$ is performed and yields a $\mathbf{R}-$matrix as a function of photon's time-of-flight $t$:
\begin{equation}
\label{time_gating}
\mathbf{R}_{\bm{\rho}\mathbf{u}}(z,t)=\int d\omega \mathbf{R}_{\bm{\rho}\mathbf{u}}(z,\omega) e^{j \omega t}.
\end{equation}
\corr{The coherence volume is defined as the ensemble of points that contribute to the singly-scattered wave-field at a given time-of-flight $t$. In an homogeneous medium of refractive index $n$, this volume is, in first approximation, a thin slice located at a depth $z_t=c_0 t /(2n)$\corr{, with $c_0$ the light velocity in vacuum}. Its thickness is governed by the light source bandwidth \corr{$\Delta \lambda$}: $\delta z_t \sim \lambda^2 / (2 n_0 \Delta \lambda)\sim 3.5$ $\mu$m.} When the focusing plane and the coherence volume coincide (Fig.~\ref{fig:image}\alex{a$_1$}), an holoscopic image of the sample, $\mathbf{I}_H$, can be obtained for each input wave-front $\mathbf{u}_\textrm{in}$ (Fig.~\ref{fig:image}a$_1$):
\begin{equation}
\label{eq3}
    I_H(\mathbf{r}_t,\mathbf{u}_\textrm{in})=R(\bm{\rho}_\textrm{out},\mathbf{u}_\textrm{in},z_t,t).
\end{equation} 
with $\mathbf{r}_t=(\bm{\rho}_\textrm{out},z_t)$. In practice, an exact matching between the focusing plane and coherence volume is difficult to obtain especially for deep imaging (\textit{i.e} low single-to-multiple scattering ratio). We will describe further how matrix imaging can provide a robust observable for this fine tuning.

Figures~\ref{fig:image}\corr{b$_1$, c$_1$ and d$_1$} display longitudinal and transverse cross-sections of the cornea obtained for a normal incident plane wave (see also Supplementary Movies 1 and 2). Although this holoscopic image can be obtained at a very high frame rate (${Fps} \sim 10^{11}$ voxels/s), it also exhibits a speckle-like feature. Indeed, \corr{multiple scattering events} taking place ahead of the \alex{coherence volume} at each time $t$ can pollute the image. \corr{Multiply-scattered waves} generate a random speckle noise without any connection with the medium reflectivity. To remove it, a naive strategy is to sum the intensity of the holoscopic images obtained for each illumination $\mathbf{u}_\textrm{in}$. \alex{Such an incoherent compound} tends to smooth out the speckle noise but the resulting \alex{image} still exhibits an extremely low contrast due to the multiple scattering background (see Supplementary Fig.~S2). To get rid of it, the single-to-multiple scattering ratio shall be increased~\cite{badon_multiple_2017}. For this purpose, a spatial filtering of multiply-scattered photons \alex{can be performed} by means of a \alex{confocal filter}. Nevertheless, this operation is extremely sensitive to the focusing quality inside the sample. A prior optimization of the focusing process is thus needed.

\vspace{5 mm}
\noindent{\textbf{Digital confocal imaging.}}

\alex{To that aim}, the dual reflection matrix is projected in the focused basis \corr{at input}. Mathematically, it simply consists in a numerical input focusing of $\mathbf{R}_{\bm{\rho}\mathbf{u}}$ \alexx{using the Fresnel propagator $\mathbf{F}_{\corr{\bm{\rho}\mathbf{u}}}$ that describes free space propagation from the MO pupil plane and the focal plane at depth $z$ (Methods\rev{, Eq.~\ref{eq23}}):
\begin{equation}
\mathbf{R}_{\bm{\rho \rho}}(z,t)=\int d\omega \mathbf{R}_{\bm{\rho }\mathbf{u}}(z,\omega) \times \mathbf{F}^{\dag}_{\corr{\bm{\rho }\mathbf{u}}}(z,\omega) e^{-j \omega t}.
\end{equation}}
where the symbol $\dag$ stands for transpose conjugate. Expressed in the focused basis, the reflection matrix $\mathbf{R}_{\bm{\rho\rho}}(z,t)$ contains the responses at each time-of-flight $t$ between virtual sensors of expected positions \corr{$(\bm{\rho}_\textrm{in},z)$} and \corr{$(\bm{\rho}_\textrm{out},z)$}.  

\begin{figure}
\includegraphics[width=\textwidth]{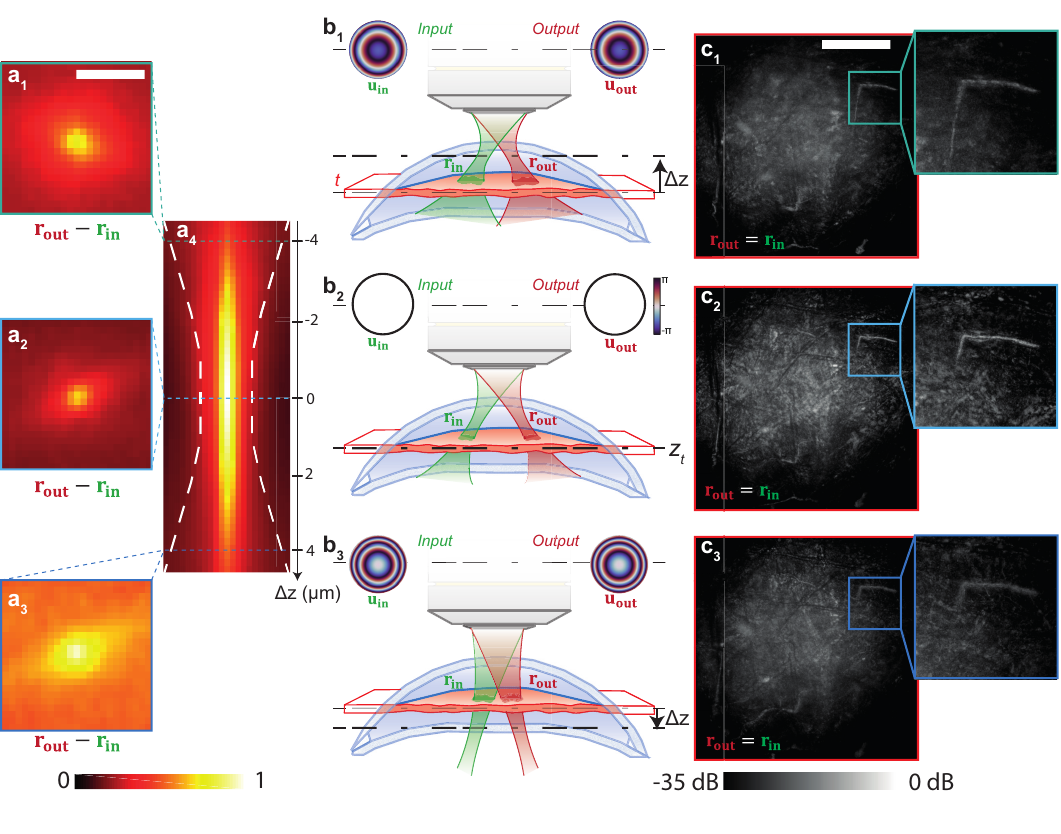}
    \caption{\textbf{Auto-focusing process guided by the reflection point spread function.} \textbf{a} Evolution of the RPSF versus the defocus distance $\Delta z$ for a fixed coherence volume. \corr{Its} transverse distribution is shown for several values of defocus (\textbf{a$_1$})-(\textbf{a$_3$}). The evolution of its radial average is displayed in panel (\textbf{a}$_4$). \corr{The color bar in a square root scale.} \textbf{b.} Relative position of the focusing plane (dash-dotted line) and coherence volume (red layer) for the different values of defocus $\Delta z$ displayed in (\textbf{a}). \textbf{c.} En-face confocal image and zoom on a nerve. In each panel, the subscripts 1, 2 and 3 stand for defocus distances $\Delta z=-4$, 0, and $+4$ $\mu$m. The considered coherence volume is located at the effective depth $z_t=140$ $\mu$m in the cornea.}
    \label{fig:refoc}
\end{figure}

The focused $\mathbf{R}$-matrix is equivalent to the time-gated reflection matrix considered in previous studies for RMI~\cite{kang_imaging_2015,badon_smart_2016,kang_high-resolution_2017,badon_distortion_2020,yoon_laser_2020,Kwon2023,Najar2023} except that we now have at our disposal a supplementary degree of freedom: \corr{the} parameter $z$ that controls the axial position of the focusing plane. A raw confocal image $\mathbf{I}_C$ can be built by considering the diagonal elements of $\mathbf{R}_{\bm{\rho \rho}}$ ($\bm{\rho}_\textrm{in}=\bm{\rho}_\textrm{out}$):
\begin{equation}
\label{eq5}
    I_C(\mathbf{r}_t,z)=R(\bm{\rho},\bm{\rho},z,t).
\end{equation} 
 Figure~\ref{fig:refoc}c shows the en-face image obtained at a given time-of-flight $t$ for different values of $z$. Qualitatively, we see that the image quality strongly depends on the relative position between the coherence volume and the focusing plane. Here the presence of a highly reflecting structure, a corneal nerve, allows us to determine the parameter $z$ that allows to match the focusing plane with the coherence volume.

\vspace{5 mm}
\noindent{\textbf{Self-portrait of the focusing process.}}

A more quantitative and robust observable is provided by the off-diagonal coefficients of $\mathbf{R}_{\bm{\rho \rho}}(z,t)$ that enable to probe the focusing quality at any voxel. More precisely, this can be done by investigating the reflection point spread function (RPSF) defined as follows:
\begin{equation}
RPSF(\Delta \bm{\rho},\bm{\rho},z,t)=  \left |R(\bm{\rho}-\Delta \bm{\rho}/2,\bm{\rho}+\Delta \bm{\rho}/2,z,t) \right |^2  ,
\end{equation}
 This quantity derived from the off-diagonal coefficients of $\mathbf{R}_{\bm{\rho \rho}}$, quantifies the focusing quality for each point $\mathbf{r}_t=(\bm{\rho},z_t)$. For a medium of random reflectivity and under a local isoplanatic assumption, its ensemble average actually scales as~\cite{lambert_reflection_2020} (Supplementary Section S5):
 \begin{equation}
 \left \langle RPSF(\Delta \bm{\rho},\bm{\rho},z,t) \right \rangle \propto |\alexx{h}_\textrm{in}|^2 \stackrel{\Delta \bm{\rho}}{\circledast}|\alexx{h}_\textrm{out}|^2(\Delta \bm{\rho}, \bm{\rho},z,t)
 \end{equation} 
 where the symbols $\langle \cdots \rangle$ and $\circledast$ stand for ensemble average and convolution product, respectively. $\alexx{h}_\textrm{in/out}(\Delta \bm{\rho},\bm{\rho}, z,t)$ is the spatial distribution of the input/output PSF along the de-scanned coordinate $\Delta \bm{\rho}$ in the coherence plane at $z_t$ when trying to focus at point $(\bm{\rho},z)$. 

The RPSF can thus provide a self-portrait of the focusing process inside the cornea. Figure \ref{fig:refoc}a shows the evolution of the laterally-averaged RPSF for a given time $t$ as a function of the parameter $z$ in the Fresnel propagator. As expected, the focusing plane and coherence volume coincide when \corr{the full width at half maximum (FWHM) $\delta \rho$ of the RPSF} is minimized (Fig.~\ref{fig:refoc}b), \textit{i.e} for a defocus distance $\Delta z=z-z_t=0$ (Fig.~\ref{fig:refoc}a$_2$). \alex{The estimated defocus is roughly constant over the whole thickness of the cornea. \alex{This} proves that the effective index of the cornea is actually very close to the refractive index $n_0$ used in our propagation model (see Supplementary Section S4).}  

\begin{figure}
    \includegraphics[width=\textwidth]{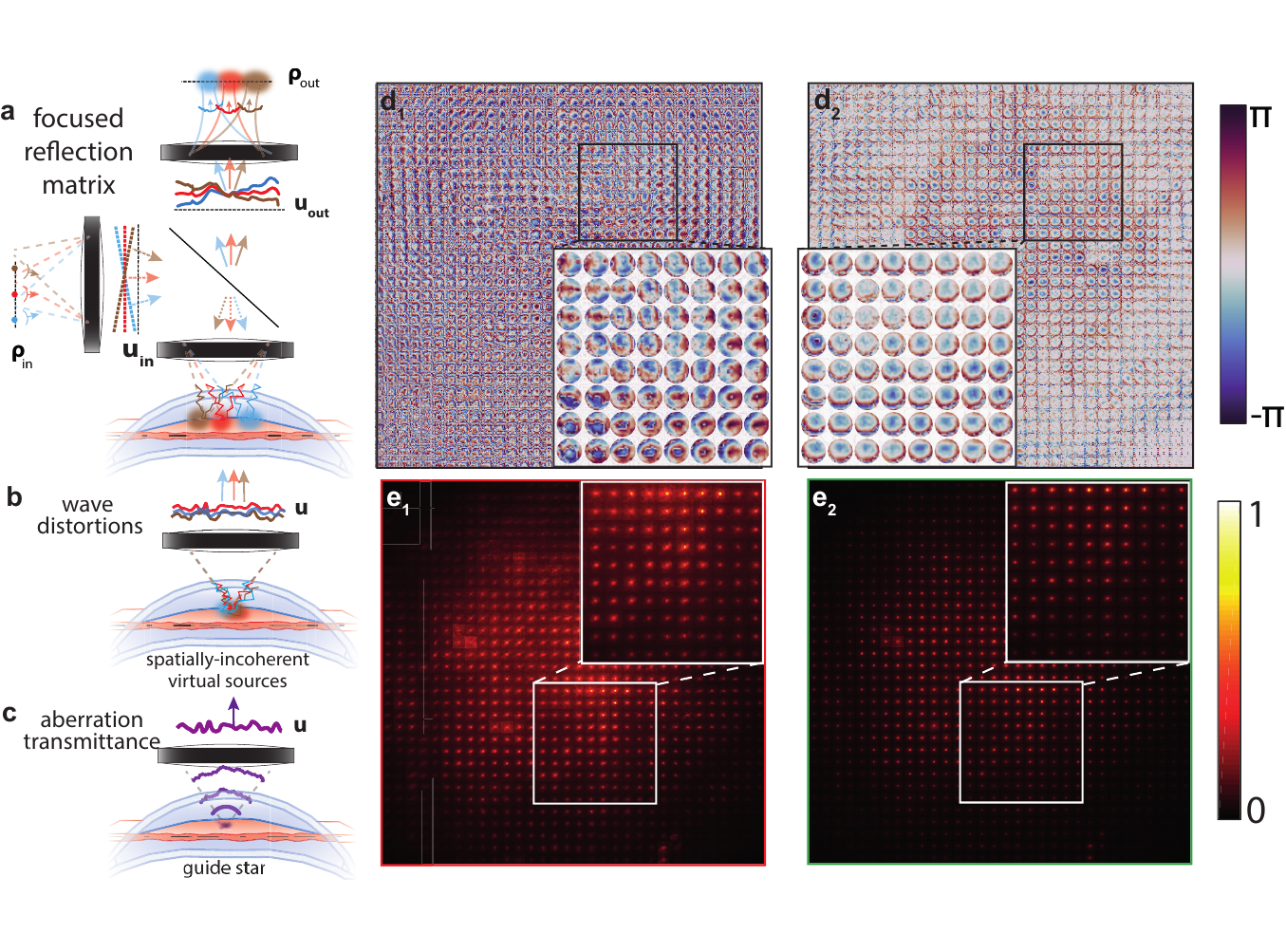}
    \caption{\corr{\textbf{Matrix Compensation of Transverse Aberrations.} \textbf{a}, The digitally-refocused $\mathbf{R}$-matrix contains the set of impulse responses $R(\bm{\rho}_\textrm{in},\bm{\rho}_\textrm{out},z_t)$ between an array of point sources $\bm{\rho}_\textrm{in}$ and detectors $\bm{\rho}_\textrm{out}$ lying in the coherence plane at depth $z_t$. \textbf{b}, Wave distortions are then isolated by projecting the optical data in the input or output pupil plane. Seen from the focal plane, this operation amounts to a de-scan of each focal spot. \textbf{c}, An iterative phase reversal process applied to those wave distortions provides a estimation of local aberration transmittances by recombining each focal spot into a virtual guide star over reduced spatial windows. \textbf{d}, The resulting maps of input (\textbf{d}$_1$) and output (\textbf{d}$_2$) aberration phase laws, ${\phi}_{\textrm{in/out}}$, are shown for depth  $z_t=140$ $\mu$m. \textbf{e}, The corresponding map of RPSFs is shown \corr{in amplitude} before (\textbf{e}$_1$) and after (\textbf{e}$_2$) the compensation of \corr{transverse aberrations} at the same depth. Each RPSF is displayed over a de-scan area of $6\times6$ $\mu$m$^2$. The color bar is in a square root scale.}} 
    \label{fig:aberration}
\end{figure}

Figures~\ref{fig:image}\corr{b$_2$, c$_2$ and d$_2$} displays longitudinal and transverse cross-sections of the confocal image obtained after tuning the coherence volume and focusing plane at any depth (see also Supplementary Movies {1 and 2}). The resolution and contrast are much better than the \alex{incoherent compound} image \alex{(Supplementary Fig.~S2)}. \corr{In particular, the axial resolution $\delta z_c$ of the digital confocal image is dictated by the high numerical aperture of the microscope objective~\cite{Sentenac2018}:} 
\corr{$\delta z_c \sim n_0 \lambda /(4 NA^2)\sim 0.5$ $\mu$m}. However, the image quality remains perfectible. Indeed, the RPSF still spreads well beyond the theoretical resolution cell \corr{with $\delta \rho \sim 2.3$ $\mu$m at $z_t=230$ $\mu$m (Fig.~\ref{fig:image}f$_1$). The Strehl ratio associated with this RPSF is extremely low ($\mathcal{S}\sim 4.5\times  10^{-3}$, Methods\rev{, Eq.~\ref{Strehl}}) and the image contrast is relatively weak ($\chi\sim 4$ dB, Methods\rev{, Eq.~\ref{Contrast}}). These imperfections are the result of \corr{transverse} aberrations originating} from the lateral fluctuations of the optical index $n(\bm{\rho},z)$ in the cornea. To demonstrate this last assertion, the transverse evolution of the focusing process can be investigated by a local assessment of the focusing quality \alex{(Methods\rev{, Eq.~\ref{RPSF}})}. A map of local RPSFs is displayed in Fig.~\ref{fig:aberration}\corr{e$_1$}. Although \corr{a prior compensation of \rev{defocus} aberrations} provides a correct focusing quality over the whole thickness of the cornea on average, the local RPSFs exhibit important fluctuations across-the field-of-view. This observation is a manifestation of the 3D distribution of the optical index $n(\mathbf{r})$ inside the cornea. This anisoplanic feature requires a local compensation of aberrations as we will see below.

\vspace{5 mm}

\noindent {\textbf{Local Compensation of Wave Distortions.}} 

\corr{By considering {the focused} reflection matrix at each depth $z_t$ (Fig.~\ref{fig:aberration}a), a local compensation of transverse aberrations can be performed  using the distortion matrix concept~\cite{badon_distortion_2020} . To that aim, the distorted component of the reflected/incident wave-front is extracted for each input/output focusing point $\bm{\rho}_{\textrm{in/out}}$ (Methods\rev{, Eq.~\ref{Distorsion}}). Wave distortions exhibit local correlations that are a manifestation of the shift-shift memory effect~\cite{judkewitz_translation_2015} characteristic of anisotropic scattering in the cornea (Fig.~\ref{fig:aberration}b). Over an isoplanatic patch, each distorted wave-field corresponds to the diffraction pattern of each virtual source ($\bm{\rho}_{\textrm{in/out}}$) modulated by the local aberration transmittance that accounts for the long-scale heterogeneities of the refractive index between the cornea surface and the coherence plane (Supplementary Section S6). Each virtual source is spatially incoherent due to the random reflectivity of the medium. The idea is to smartly combine each of them to synthesize a coherent guide star and estimate an aberration phase law independently from the sample reflectivity.  In practice, this is done by applying an iterative phase reversal \rev{(IPR)} process~\cite{Bureau2023} around each voxel $(\bm{\rho},z_t)$ (Methods\rev{, Eq.~\ref{IPR}}). The result is an estimation of the local aberration transmittance in the pupil plane~\cite{Najar2023}(Fig.~\ref{fig:aberration}c). The same operation can be repeated for each voxel of the medium. The set of extracted aberration phase laws form the input and output aberration phase matrices, $\bm{\Phi}_{\textrm{in/out}}(z_t)=[{\phi}_{\textrm{in/out}}(\mathbf{u}_\textrm{in/out},\bm{\rho},z_t)]$, between the pupil plane ($\mathbf{u}_\textrm{in/out}$) and the medium voxels $(\bm{\rho},z_t)$~\cite{Bureau2023}.}

\corr{A crucial parameter is the spatial extent $L$ of the spatial window over which correlations between wave distortions shall be investigated  (Methods\rev{, Eq.~\ref{corr_out}}). This choice actually governs the transverse resolution of $\bm{\Phi}_{\textrm{in/out}}(z_t)$ and is subject to the following dilemma: On the one hand, the spatial window should be as small as possible to grasp the rapid variations of the aberration phase laws across the field of view; on the other hand, these areas should be large enough to encompass a sufficient number $N_L=(L/\delta \rho_f)^2$ of independent realizations of disorder, with $\delta \rho_f \sim \lambda/(2NA)$ the standard diffraction-limited resolution.  Indeed, the bias of our aberration phase estimator is inversely proportional to $N_L$~\cite{Bureau2023}. In the present case, the optimal value for $L$ has been found to be $18.6$ $\mu$m.} 

The result of the aberration correction process is displayed in Fig.~\ref{fig:aberration}\corr{d} at {$z_t=140$} $\mu$m. Strikingly, the estimated aberration laws exhibit strong phase fluctuations and vary quickly between neighboring windows. This complex feature has two origins: (\textit{i}) the lateral fluctuations exhibited by the optical index inside the cornea; (\textit{ii}) the imperfections of the imaging system. The latter component accounts for the difference observed between the input and output aberration transmittances \alex{(Supplementary Section S4)}. In fact, the input aberration phase law accumulates not only the input aberrations of the sample-arm but also those of the reference arm. The sample-induced aberrations can be investigated independently from the imperfections of the experimental set up by considering the output aberration phase matrix $\bm{\Phi}_{\textrm{out}}$ \corr{(Fig.~\ref{fig:aberration}d$_2$)}. \corr{Despite the prior compensation of the mean defocus by minimizing the RPSF width, the aberration phase laws exhibit a residual defocus} that varies across the field-of-view due to \corr{the} lateral variations of the optical index. Local shifts of the pupil function are also observed and result from a local curvature of the coherence \corr{surface} with respect to the focusing plane. 

\corr{The extracted aberration phase laws can be used to estimate the transmission matrices, $\mathbf{G}_{\textrm{in/out}}(z_t)=[G_\textrm{in/out}(\bm{\rho}_\textrm{in/out},\bm{\rho}',z_t)]$, containing the impulse responses $G_\textrm{in/out}(\bm{\rho}_\textrm{in/out},\bm{\rho}',z_t)$ between the image voxels ($\bm{\rho}_\textrm{in/out}$) and the sample mapped by the vector $\bm{\rho}'$ at each depth $z_t$ (Methods, Eq.~\ref{Tmatrix}).} \corr{Aberrations are then compensated by applying the phase conjugate of the {transmission} matrices~\cite{Popoff2010} at {the} input and output \corr{of the focused $\mathbf{R}-$matrix}} (Fig.~\ref{fig:image}a$_3$), such that:
\begin{equation}
\label{deconvolution}
\mathbf{R}^{\corr{''}}_{\bm{\rho \rho}}(z_t)=\alexx{\mathbf{G}_{\textrm{out}}^\dag} (z_t) \times \mathbf{R}_{\bm{\rho \rho}}(z_t) \times \alexx{\mathbf{G}_{\textrm{in}}^*} (z_t) 
\end{equation}
The final image of the sample can be obtained by considering the diagonal elements of the corrected matrix $\mathbf{R}^{\corr{''}}_{\bm{\rho \rho}}$:
\begin{equation}
\label{eq9}
I_M(\mathbf{r}_t)=R^{\corr{''}}(\bm{\rho},\bm{\rho},z_t).
\end{equation} 
Figures~\ref{fig:image}\corr{b$_3$, c$_3$ and d$_3$} display longitudinal and transverse cross-sections of the cornea obtained by RMI (see also Supplementary Movies 1, 2 and 3). The comparison with the confocal image [Figs.~\ref{fig:image}\corr{b$_2$, c$_2$ and d$_2$}] shows a clear gain in contrast \corr{and resolution}. This \corr{drastic improvement of the image} can be quantified by examining the RPSF. While, at the previous step, the confocal peak exhibits a spreading well beyond the diffraction limit and a background at depth due to forward multiple scattering events (Figs.~\ref{fig:image}e$_1$,f$_1$), RMI compensates for these two issues and leads to an almost ideal RPSF (Figs.~\ref{fig:image}e$_2$,f$_2$) over the whole field-of-view (Fig.~\ref{fig:aberration}\corr{e$_2$}). \corr{The image improvement can be assessed by the RPSF (Fig.~\ref{fig:image}f$_3$) whose Strehl ratio $\mathcal{S}$ is increased by a factor 7 compared to its original value (Fig.~\ref{fig:image}f$_2$). At the same depth $z_t=230$ $\mu$m, the contrast $\chi$ is increased by a factor 11 and the resolution $\delta \rho $ is enhanced by a factor four (Fig.~\ref{fig:image}f$_3$), thus nearly reaching the diffraction-limited value $\rho_f\sim 500$ nm.} 

The obtained three-dimensional image highlights several crucial features of the cornea: its lamellar structure induced by the collagen fibrils (Fig.~\ref{fig:image}b$_3$); (\textit{ii}) the complex network of nerves that covers the cornea; (\textit{iii}) characteristic structures of the cornea such as keratocytes and; (\textit{iv}) stromal striae whose presence is an indicator of keratoconus~\cite{Grieve2017}. Such a high-resolution image can thus be of particular importance for bio-medical diagnosis, given the high frame rate of our device. Of course, RMI is not limited to the cornea but can be also applied to the deep inspection of retina, skin or arteries, tissues whose structures are already monitored by OCT but, until now, limited by a modest penetration depth. 

\vspace{5 mm}

\noindent {\rev{\textbf{Addressing Forward Multiple Scattering.}}}

\begin{figure}
    \includegraphics[width=\textwidth]{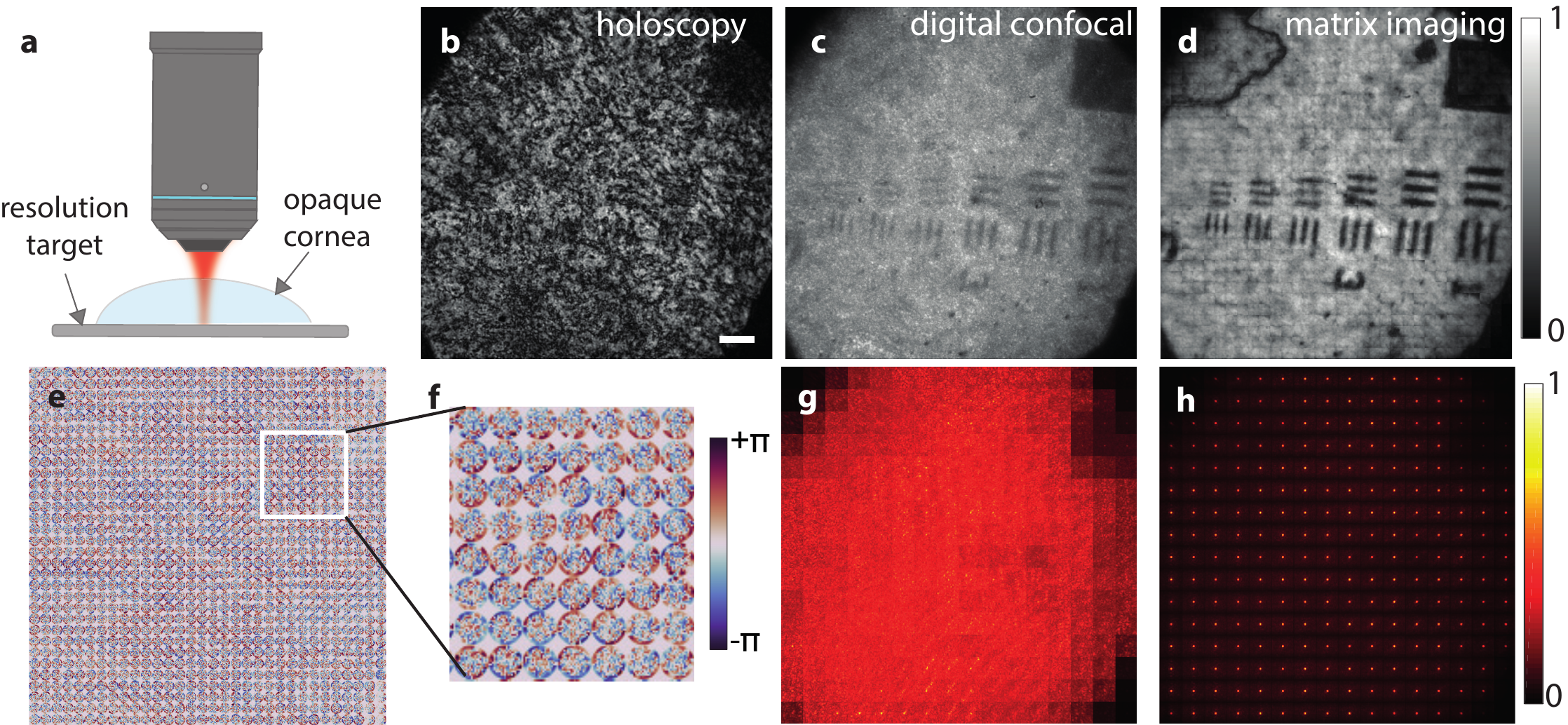}
    \caption{\rev{\textbf{Imaging a resolution target through the cornea.} \textbf{a}, Experimental configuration. \textbf{b}, Holoscopy [scale bar: 20$\mu$m]. \textbf{c}, Digital confocal image. \textbf{d}, Matrix image based on the IPR algorithm. \textbf{e}, Maps of the aberration phase laws estimated from the pupil plane.  \textbf{f}, Zoom on the white rectangle displayed in panel e. \textbf{g}, Maps of RPSFs obtained after the digital focusing process. \textbf{h}, Maps of RSPFs after aberration matrix compensation. Each RPSF is displayed over
a de-scan area of 6.3 $\times$ 6.3 $\mu$m$^2$.} } 
    \label{fig:mire}
\end{figure}

\rev{In that respect, 
Fig.~\ref{fig:mire} shows that RMI can also succeed in a stronger (forward) multiple scattering regime, by imaging a resolution target through a more opaque region of the cornea, closer to the iris (Fig.~\ref{fig:mire}a, see Methods). Figures~\ref{fig:mire}b and c display the images obtained for a single illumination ($\mathbf{u}_\textrm{in}=\mathbf{0}$, Eq.~\ref{eq3}) and after the coherent compound of $N_{\textrm{in}}=325$ illuminations (Eq.~\ref{eq5}). While the holoscopic image exhibits a random speckle due to multiple scattering (Fig.~\ref{fig:mire}b), most of the patterns of the resolution target are revealed by the digital confocal image (Fig.~\ref{fig:mire}c). However, it is poorly contrasted ($\chi\sim 0.9$ dB). This is confirmed by the blurred feature of the RPSFs characteristic of multiple scattering, which highlights the cornea turbidity (Fig.~\ref{fig:mire}g). The associated Strehl ratio is extremely low: $\mathcal{S}\sim 10^{-3}$. To tackle multiple scattering, a local analysis of wave distortions is then performed (Methods\rev{, Eq.~\ref{corr_out}}). The corresponding aberration phase laws are displayed in Fig.~\ref{fig:mire}e,f. They exhibit a complex feature characteristic of forward multiple scattering with a broad spatial frequency content and a short-scale memory effect~\cite{Najar2023}. The phase conjugate of the associated transmission matrix (Eq.~\ref{deconvolution}) provides the final image (Eq.~\ref{eq9}) displayed in Fig.~\ref{fig:mire}d. The high contrast of the image demonstrates the benefit of RMI. The comparison between original and final RPSFs confirms the drastic improvement of the focusing quality (Figs.~\ref{fig:mire}g,h).  The transverse resolution almost reaches the confocal resolution $\delta \rho_c$: {$\delta \rho \sim 0.35$ $\mu$m}.}


The ability of RMI in overcoming high-order aberrations and multiple scattering thus constitutes a paradigm shift for deep optical microscopy. \corr{In its present form, the penetration depth of RMI remains of the order of $\ell_t$, the typical depth beyond which the memory effect vanishes~\cite{Osnabrugge2017}. However, a multi-scale~\cite{Najar2023} and/or multi-conjugate~\cite{Kang2023} compensation of wave distortions can actually address even more complex scattering trajectories associated with extremely small isoplanatic patches (a few $\mu$m). Moreover, the multi-spectral reflection matrix gives access to temporal degrees of freedom that can be exploited for tailoring complex spatio-temporal focusing laws~\cite{Lee2023} required to overcome the diffusive limit. The mapping of the refractive index will also be an important step to build  accurate focusing laws inside the medium~\cite{Chen2020}. As shown by quantitative phase imaging of thin biological samples, this physical parameter is also a quantitative marker for biology. Mapping the refractive index in 3D and in an epi-detection geometry will pave the way towards a \alex{quantitative} imaging of biological tissues.}
\vspace{5 mm}

\noindent {\large \textbf{Discussion}}

In contrast with previous works that considered the reflection matrix at a single frequency~\citep{popoff_exploiting_2011} or time-of-flight~\citep{kang_imaging_2015,badon_smart_2016,yoon_laser_2020}, the measurement of a polychromatic reflection matrix~\cite{Zhang2023} allowed us to realize in post-processing: (i) a 3D confocal image of the sample reflectivity on millimetric volumes (0.1 mm$^3$ = $\corr{5}\times 10^9$ voxels) in an ultra-fast acquisition time (\corr{1.4} s) ; (ii) a local compensation of aberrations which usually prevent deep imaging. 

\corr{The required number of input wave-fronts depends on the aberration level and scales as the number of resolution cells covered by the RPSF~\cite{Bureau2023}. If this condition is not met, the diffuse halo extends beyond the RPSF extension \rev{(Fig.~\ref{fig:mire}g)} and suppression of multiple scattering is then less efficient than it would be in conventional confocal imaging \rev{(see the residual bakground in Fig.~\ref{fig:mire}h)}. In practice, a compromise shall therefore be made between the frame rate and the image quality. Note also that a great advantage of the matrix approach with respect to conventional confocal imaging lies in its ability of local aberration compensation in post-processing.}
\corr{Moreover, a spectral measurement and spatial multiplexing of the wave-field provides a drastic enhancement of the signal-to-noise ratio by a factor $N_{\omega}\times N_{\textrm{in}}$, \textit{i.e} the number of spatio-temporal degrees of freedom provided by our illumination sequence~\cite{mertz_introduction_2019}.} 


An issue we have not considered in this paper is \alex{medium motion} during the acquisition of the reflection matrix. Of course, the assumption of a static medium is everything but true especially for in-vivo applications~\citep{Jang2014}. To cope with the dynamic features of the medium, two strategies can be followed. The first one is to limit the measurement time of the $\mathbf{R}$-matrix at its minimum, as allowed by our device using a few illuminations. The second one is to develop algorithms that consider \alex{medium motion} during the measurement of $\mathbf{R}$~\cite{Kang2023}.  Interestingly, temporal fluctuations of the medium's reflectivity and refractive index can provide a key information for probing the multi-cellular dynamics in optical microscopy~\cite{Apelian2016,Scholler2020}. 

Just as the concept of plane-wave imaging~\cite{montaldo_coherent_2009} revolutionized the field of ultrasound~\cite{mace_functional_2011,errico_ultrafast_2015} by providing an unprecedented frame rate, \corr{our device will constitute in a near future an ideal tool for probing the 3D dynamics of tissues at a much smaller scale~\cite{Scholler2020,Muenter2020}. In medical imaging, an increase of the frame rate by one to two orders of magnitude is necessary for ophtalmic applications due to eye movements (10 Hz~\cite{Mazlin2022}). It is also important in dermatology since standard OCT devices remain extremely slow to acquire 3D skin images (0.04 Hz~\cite{Latriglia2023}). At last, our imaging apparatus is an ideal tool for the monitoring of cell development in 3D whether it be for embryology or pharmacology with the fast development of organoids~\cite{Harrison2021}. For all these applications, the 3D imaging capability at a high frame rate will be particularly decisive.}

\newpage

\noindent {\Large \textbf{Methods}} 

\noindent {\textbf{Experimental components.}} 

The following components were used in the experimental setup (Fig.~\ref{fig:setup}): A swept laser source (800-875 nm; Superlum-850 HP), one galvanometer (Thorlabs, LSKGG4), one scan lens $L_1$ ($f_1=$ 110 mm), two immersion objective lenses (40$\times$; NA, 0.8; \Paul{Nikon}), an imaging lens $L_2$ ($f_2=250$ mm) and an ultrafast camera (25~kHz; Phantom-v2640). 

\vspace{5 mm}

\noindent {\textbf{Sample preparation.}}  

In the  presented experiment, the corneal sample was fixed with paraformaldehyde (4\% concentration). \corr{Note that this substance tends to enhance scattering while absorption remains unchanged~\cite{Wood2011}.}
 
\vspace{5 mm}

\noindent {\textbf{Sampling of input and output wave-fields.}} 

The dimension of the input pupil is $D_{in} \times D_{in} = 9 \times 9$ mm; the spatial sampling of input wave-fields is \alex{$\delta u_\textrm{in}=\Paul{600 \mu}$m}. Given the magnification of the output lens system (MO, L$_2$) system MO1 and the inter-pixel distance of the camera ($\delta s_\textrm{out}=12$ $\mu$m), the output wave-field is sampled at a resolution close to $\lambda/(4NA)$\corr{, the theoretical achievable resolution for a confocal image~\cite{Sentenac2018}}: $\delta{\rho}_{\corr{c}} = 290$ nm.

\vspace{5 mm}

\noindent {\textbf{Data acquisition and GPU processing.}} 

All the interferograms of the acquisition sequence are recorded by the camera in 1.4 s and stored in its internal memory. Then, the whole data set (75 \corr{Gb}) is transferred to the computer in \alex{5 min}. The numerical post-processing of the reflection matrix is performed by GPU (NVIDIA TITAN RTX) and takes \alex{3.6 s} per input wave-fronts. For the data set considered in this paper, all the focusing and aberration correction algorithms are performed in 1 hour.

\vspace{5 mm}
\noindent {\textbf{On-axis holography.}}

For each input wave-field, the interferogram recorded by the camera can be expressed as follows:
\begin{equation}
	I(\mathbf{s}_\textrm{out},\mathbf{u}_\textrm{in},\omega) = \left| \alexx{E}(\mathbf{s}_\textrm{out},\alexx{\mathbf{u}_\textrm{in},}\omega) + \alexx{E_\textrm{ref}}(\mathbf{s}_\textrm{out},\mathbf{u}_\textrm{in},\omega) \right|^2
\end{equation}
with $E$ and $E_\textrm{ref}$, the wave-fields reflected by the sample and reference arms. Then a Fourier transform in the frequency domain is performed. The resulting intensity can be written as follows:
\begin{align}
	I(\mathbf{s}_\textrm{out},\mathbf{u}_\textrm{in},t)& =  E(\mathbf{s}_\textrm{out},\uin,t)\stackrel{t}{\circledast} E^*(\mathbf{s}_\textrm{out},\uin,-t) \label{sample}  \\
 & + E_\textrm{ref}(\mathbf{s}_\textrm{out},\uin,-t)\stackrel{t}{\circledast} E_\textrm{ref}^*(\mathbf{s}_\textrm{out},\uin,-t) \label{reference} \\ 
 &+ E_\textrm{ref}(\mathbf{s}_\textrm{out},\uin,t)\stackrel{t}{\circledast} E^*(\mathbf{s}_\textrm{out},\uin,-t) \label{anticausal}\\
 & + E(\mathbf{s}_\textrm{out},\uin,t) \stackrel{t}{\circledast}E^{*}_\textrm{ref}(\mathbf{s}_\textrm{out},\uin,-t)
 \label{causal}
\end{align}
The two first terms (Eqs.~\ref{sample} and \ref{reference}) correspond to the self-interference of each arm with itself. Both contributions emerge at an optical depth close to zero ($t=0$). The two last terms correspond to the anti-causal (Eq.~\ref{anticausal}) and causal (Eq.~\ref{causal}) components of the interference between the two arms. By applying a Heavyside filter to $I(\mathbf{s}_\textrm{out},\mathbf{u}_\textrm{in},t)$ along the time dimension, one can isolate the causal contribution (Eq.~\ref{causal}).  An inverse Fourier transform then yields the distorted wave-field:
\begin{equation}
D(\mathbf{s}_\textrm{out},\mathbf{u}_\textrm{in},\omega)=E(\mathbf{s}_\textrm{out},\uin,\omega)E^*_\textrm{ref}(\mathbf{s}_\textrm{out},\uin,\omega). 
\end{equation}
If aberrations in the reference arm are neglected \alexx{(Supplementary Section S3)}, the reference wave-field is a replica of the incident wave-field,
$$E_\textrm{ref}(\mathbf{s}_\textrm{out},\uin,\omega)=\exp \left (i \frac{2\pi}{\lambda f} \mathbf{u}_\textrm{in}.\mathbf{s}_\textrm{out} \right ).$$ The multi-spectral reflection matrix is thus extracted using the following relation:
\begin{equation}
	\mathbf{R}(\mathbf{s}_\textrm{out},\uin,\omega) = \mathbf{D}(\mathbf{s}_\textrm{out},\uin,\omega) \exp \left (-i \frac{2\pi}{\lambda f} \mathbf{u}_\textrm{in}.\mathbf{s}_\textrm{out} \right ).
\end{equation}

\vspace{5 mm}

\noindent {\textbf{Fresnel operators}} 

\alexx{The numerical focusing process is performed by means of Fresnel propagators. For this purpose, the multi-spectral reflection matrix should be first projected in the output pupil plane ($\mathbf{u}_\textrm{out}$) by a simple 2D spatial Fourier transform: 
\begin{equation}
\label{projpupil}
\mathbf{R}_{\mathbf{uu}}(\omega)=\mathbf{P}_{\mathbf{us}}^{\alex{*}}(\omega) \times \mathbf{R}_{\mathbf{su}}(\omega)
\end{equation}
where $\mathbf{P}_{\mathbf{us}}\corr{(\omega)}=[P(\mathbf{u},\mathbf{s},\corr{\omega})]$ is the Fourier transform operator:
\begin{equation}
P(\mathbf{u},\mathbf{s},\omega)=e^{-j\alex{\frac{\omega}{{c}_0}\frac{\mathbf{u}\cdot \mathbf{s}}{f}}}
\end{equation}
with $f$, the focal length of the MOs and $c_0$ the vacuum light velocity. A Fresnel phase law is then applied at the output of $\mathbf{R}_{\mathbf{uu}}(\omega)$ to numerically shift the focal plane, originally located in the middle of the sample ($z=0$) to any depth $z$:
\begin{equation}
\label{fresnel1}
\mathbf{R}_{\bm{\rho}\mathbf{u}}(\alex{z},\omega)=[\mathbf{P}^{\top}_{\mathbf{u}\bm{\rho}}(\omega) \circ \bm{\mathcal{F}}_{\mathbf{u}}(z,\omega) ] \times \mathbf{R}_{\mathbf{u}\mathbf{u}}(\omega)
\end{equation}
where the symbol $\circ$ accounts for the Hadamard (term-by-term) product. Each column vector $\bm{\mathcal{F}}_{\corr{\mathbf{u}}}(z,\omega)$ is a phase mask that accounts for the propagation of each plane wave of transverse wave vector \alex{$\mathbf{k}_{||}=\omega\mathbf{u}/({c_0}f)$} over a thickness $z$ of an homogeneous medium of refractive index $n_0$:
\begin{equation}
\label{fresnel2}
	\mathcal{F}(\mathbf{u},z,\omega) = e^{-j\left (\frac{n_0 \omega}{c_0}-k_z \right )z} \alex{\mathcal{O}{(\mathbf{u})}}
\end{equation}
with
\begin{equation}
	k_z = \frac{\omega}{c_0}\sqrt{n_0 ^2-\frac{||\mathbf{u}||^2}{f^2}},
\end{equation}
the longitudinal component of the wave \alex{vector, and $\mathcal{O}{(\mathbf{u})}$, the finite pupil support: $\mathcal{O}{(\mathbf{u})}=1$ for $||\mathbf{u}||<f NA$ and zero elsewhere.} Each reflection matrix $\mathbf{R}_{\bm{\rho}\mathbf{u}}(z,\omega)=[R(\bm{\rho}_\textrm{out},\mathbf{u}_\textrm{in},z ,\omega)]$ connects each output virtual focusing point $\mathbf{r}_\textrm{out}=(\bm{\rho}_\textrm{out},z)$ to each input illumination $\mathbf{u}_\textrm{in}$ at frequency $\omega$.} \alexx{The combination of Eqs.~\ref{projpupil}, \ref{fresnel1} and \ref{fresnel2} leads us to define the Fresnel operator $\mathbf{F}_{\bm{\rho}\mathbf{s}}(z,\omega)=[F(\corr{\bm{\rho}_\textrm{out},\mathbf{s}_\textrm{out}},z,\omega)]$ that enables the projection of the optical data from the camera sensors ($\mathbf{s}_\textrm{out}$) to any focal plane ($\bm{\rho}_\textrm{out}$) at expected depth $z$ (Eq.~\ref{dualR}):
\begin{equation}
\label{eq22}
\mathbf{F}_{\corr{\bm{\rho}\mathbf{s}}}(z,\omega)=[\mathbf{P}^{\top}_{\mathbf{u}\bm{\rho}}(\omega) \circ \bm{\mathcal{F}}_{\mathbf{u}}(z,\omega) ] \times \mathbf{P}_{\mathbf{us}}^{\alex{*}}(\omega)
\end{equation}}

\alexx{The projection between the plane wave illumination basis ($\mathbf{u}_\textrm{in}$) and the focused basis (\corr{$\bm{\rho}_\textrm{in}$}) can also performed by means of a Fresnel propagator $\mathbf{F}_{\bm{\rho}\mathbf{u}}$ this time defined between the pupil plane and the each focal plane identified by their depth $z$:
\begin{equation}
\label{eq23}
\mathbf{F}_{\bm{\rho}\mathbf{u}}(z,\omega)=\mathbf{P}^{\corr{\dag}}_{\mathbf{u}\bm{\rho}}(\omega) \circ \bm{\mathcal{F}}_{\mathbf{u}}(z,\omega)
\end{equation}}

\vspace{5 mm}

\noindent {\textbf{Local estimation of focusing quality}} 

To probe the local RPSF, the field-of-view is divided at each effective depth $z_t$ into regions that are defined by their central midpoint $\bm{\rho}_\textrm{p}=(x_\textrm{p},y_\textrm{p})$ and their spatial extension $L$. A local average of the back-scattered intensity can then be performed in each region:
\begin{equation}
\label{RPSF}
  RPSF_l(\Delta \bm{\rho},\bm{\rho}_p,z,z_t)=\langle |R(\bm{\rho}+\Delta \bm{\rho}/2,\bm{\rho}-\Delta \bm{\rho}/2,z,t) |^2 W_{L}(\bm{\rho}- \bm{\rho}_p) \rangle_{\bm{\rho}}
\end{equation}
where the symbol $\langle \cdots \rangle_{m}$ stands for an average over the variable $m$ in subscript.  $W_{L}(\bm{\rho}- \bm{\rho}_p) = 1$ for $|x - x_\mathrm{p} |<L/2$ and $|y - y_\mathrm{p} |<L/2$, and zero otherwise. In this paper, a spatial window of size $L=18.6$ $\mu$m has been used to smooth out fluctuations due to the sample inhomogeneous reflectivity~\cite{lambert_reflection_2020}. 

\vspace{5 mm}

\noindent \corr{\noindent {\textbf{Strehl ratio and contrast}} }

\corr{On the one hand, a quantification of the focusing quality can be peformed with the RPSF by computing the associated Strehl ratio $\mathcal{S}$. In reflection, this quantity can actually be defined as the ratio between the confocal energy and the overall backscattered energy, such that:
\begin{equation}
\label{Strehl}
\mathcal{S}=\frac{RSPSF(\Delta \bm{\rho}=\mathbf{0})}{\sum_{\Delta \bm{\rho}}RSPSF(\Delta \bm{\rho})}.
\end{equation}
On the other hand, the image contrast $\chi$ can be evaluated by considering the ratio between the confocal peak of the RPSF and its diffuse background:
\begin{equation}
\label{Contrast}
\chi=\frac{RPSF(\Delta \bm{\rho}=\mathbf{0})}{\langle RSPSF(\Delta \bm{\rho})\rangle_{||\Delta \bm{\rho}||>2\delta \rho}}.
\end{equation}
where the diffuse background is here estimated by averaging the RPSF outside the confocal peak, \textit{i.e} arbitrarily beyond $2 \delta \rho$.}

\vspace{5 mm}

\noindent \alex{\textbf{Local compensation of wave-distortions}}

The starting point is the time-gated reflection matrix $\mathbf{R}_{\bm{\rho \rho}}(z_t)$, obtained after tuning the focusing plane and coherence volume at each echo time $t$. The first step is a projection of $\mathbf{R}_{\bm{\rho \rho}}(z_t)$ in the pupil plane at input via a numerical Fourier transform:
\begin{equation}
\mathbf{R}_{ \bm{\rho} \mathbf{u} }(z_t)= \mathbf{R}_{\bm{\rho \rho}}(z_t)   \corr{\times \mathbf{P}_{\mathbf{u}\bm{\rho}}^{\top}} \corr{(\omega_c)}
\end{equation}
\corr{with $\omega_c$, the central frequency.} An input distortion matrix is then built by performing a element-wise product between $\mathbf{R}_{ \mathbf{u} \bm{\rho}}(z_t)$ and the phase conjugate reference matrix $\corr{\mathbf{P}}_{\mathbf{u} \bm{\rho}}(\omega_c)$ that would be obtained in absence of aberrations~\cite{badon_distortion_2020} (Supplementary Section S6):
\begin{equation}
\label{Distorsion}
\mathbf{D}_{  \bm{\rho}\mathbf{u}}(z_t)= \mathbf{R}_{ \bm{\rho}\mathbf{u} }(z_t) \circ \alexx{\mathbf{P}}_{\mathbf{u}\bm{\rho}}^{\dag}\corr{(\omega_c)}
\end{equation}
A local correlation matrix $\mathbf{C}_{\textrm{in}}$ of wave distortions is then built around each point $\rp=(\bm{\rho}_\textrm{p},z_t)$ of the field-of-view (Supplementary Section S7). Its coefficients write:
\begin{equation}
\label{corr_in}
    C_\textrm{in} (\mathbf{u}_\textrm{in}, \mathbf{u}'_\textrm{in},\rp)= \left \langle {D}({\bm{\rho}_\textrm{out},\mathbf{u}_\textrm{in},z_t}){D}(\alex{\bm{\rho}_\textrm{out},\mathbf{u}'_\textrm{in},z_t}) W_{{L}}(\alex{\bm{\rho}_\textrm{out} - \bm{\rho}_\textrm{p}})\right \rangle _{\alex{\bm{\rho}_\textrm{out}}}
\end{equation}
Iterative phase reversal (see further) is then applied to each correlation matrix $\mathbf{C}_\textrm{in}(\rp)$~\cite{Najar2023} (Supplementary Section S8).  The resulting input phase laws, ${\bm{\Phi }}_{\textrm{in}}(z_t) =[{{\phi }}_{\textrm{in}}(\mathbf{u}_\textrm{in},\bm{\rho}_p,z_t)] $, are used to compensate for the wave distortions undergone by the incident wave-fronts:
\begin{equation}
\label{input_correction}
    {\mathbf{R}}'_{\bm{\rho \rho}} (z_t)=  \left \lbrace \mathbf{R}_{\bm{\rho}\mathbf{u} }(z_t)\circ \exp \left [ -i {\bm{\Phi }}_{\textrm{in}}(z_t) \right]   \right \rbrace \times \alexx{\mathbf{P}_{\mathbf{u}\bm{\rho}}^*}
\end{equation}
The corrected matrix ${\mathbf{R}}'_{\bm{\rho \rho}}$ is only intermediate since phase distortions undergone by the reflected wave-fronts remain to be corrected.

\alex{To that aim, $\mathbf{R}'_{\bm{\rho \rho}}(z_t)$} is now projected in the pupil plane at output:
\begin{equation}
\label{change}
    \mathbf{R}'_{\mathbf{u}\bm{\rho}}(z_t)= \alexx{\mathbf{P}_{\mathbf{u}\bm{\rho}}}\times \mathbf{R}'_{\bm{\rho}\bm{\rho}}(z_t).
\end{equation}
An output distortion matrix is then built:
\begin{equation}
 \mathbf{D}_{\mathbf{u}\bm{\rho}}(z_t)=  \alexx{\mathbf{P}_{\mathbf{u}\bm{\rho}}^*}\circ \mathbf{R}'_{\mathbf{u}\bm{\rho}}(z_t)
\end{equation}
From $\mathbf{D}_{\mathbf{u}\bm{\rho}}$, one can build a correlation matrix $\mathbf{C}_\textrm{out}$ for each point $\rp$:
\begin{equation}
\label{corr_out}
    C_\textrm{out} (\mathbf{u}_\textrm{out}, \mathbf{u}'_\textrm{out},\rp)= \left \langle D(\mathbf{u}_\textrm{out},\bm{\rho}_\textrm{in} ,{z_t}){D}_\textrm{out}^{*}(\mathbf{u}'_\textrm{out},\bm{\rho}_{\textrm{in} },{z_t}) W_{L}(\bm{\rho}_\textrm{in} - \bm{\rho}_\textrm{p})\right \rangle _{\bm{\rho}_\textrm{in}}
\end{equation}
The IPR algorithm described further is then applied to each matrix $\mathbf{C}_\textrm{out}(\rp)$. The resulting output phase laws, ${\bm{\Phi}}_{\textrm{out}}(z_t)=[{{\phi}}_\textrm{out}(\mathbf{u}_\textrm{out},\bm{\rho_p},z_t)]$, are leveraged to compensate for the residual wave distortions undergone by the reflected wave-fronts:
\begin{equation}
\label{output_correction}
    {\mathbf{R}}''_{\bm{\rho \rho}} (z_t)= \alexx{\mathbf{P}}_{\mathbf{u}\bm{\rho}}^\dag \times  \left \lbrace \exp \left [ -i {\bm{\Phi }}_{\textrm{out}} (z_t) \right ]   \circ  \mathbf{R}'_{\mathbf{u}\bm{\rho} }(z_t)\right \rbrace
\end{equation}

\noindent {\textbf{Iterative phase reversal algorithm.}} 

The IPR algorithm is a computational process that provides an estimator of the phase of the transmittance that links each point $\mathbf{u}$ of the pupil plane with each voxel $\rp=(\bm{\rho}_p,z_t)$ of the cornea volume~\cite{Najar2023}. To that aim, the correlation matrix $\mathbf{C}$ computed over the spatial window $W_{L}$ centered around a given point $(\bm{\rho}_p,z_t)$ is considered {(Eqs.~\ref{corr_in} and ~\ref{corr_out})}.  Mathematically, the algorithm is based on the following recursive relation:
\begin{equation}
\label{IPR}
{\bm{\Phi}}^{(n)}_\textrm{in/out} (\bm{\rho}_p,z_t) = \mathrm{arg}\left\{  \mathbf{C}_{\corr{\textrm{in/out}}}(\bm{\rho}_p,z_t) \times \exp \left [ i {\bm{\Phi}}^{(n-1)}_\textrm{in/out}(\bm{\rho}_p,z_t) \right ]  \right\}
\end{equation}
where {${\bm{\Phi}}_\textrm{in/out}^{(n)}$} is the estimator of the transmittance phase at the $n^\textrm{th}$ iteration of the phase reversal process. {$ {\bm{\Phi}}_\textrm{in/out}^{(0)}$} is an arbitrary wave-front that initiates the process (typically a flat phase law) and ${{\bm{\phi}}}_\textrm{in/out}= \lim_{n\to\infty} {\bm{\phi}}_\textrm{in/out}^{(n)}$ is the result of IPR. \\

\noindent {\textbf{Transmission matrices.}} 

\alexx{The transmission matrices $\mathbf{G}_\textrm{in/out}$ used to deconvolve the focused $\mathbf{R}$-matrix (Eq.~\ref{deconvolution}) can be deduced from the estimated aberration phase laws  ${{\bm{\Phi}}}_\textrm{in/out}$ as follows (Eqs.~\ref{input_correction} and \ref{output_correction}): 
\begin{equation}
\label{Tmatrix}
\mathbf{G}_\textrm{in/out}(z_t)= \left \lbrace \mathbf{P}_{\mathbf{u}\bm{\rho}}^{\corr{T}} \circ \exp \left [ i \bm{\Phi}_{\textrm{in/out}}(z_t) \right \rbrace  \right ] \times \mathbf{P}_{\mathbf{u}\bm{\rho}}^{\corr{*}}.
\end{equation}}\\

\noindent \rev{\textbf{Resolution target experiment.}} \\

\rev{A resolution target is placed behind the cornea in the focal plane of the microscope objective (NA=0.8). The reflection matrix $\mathbf{R}$ is measured with $N_{\textrm{in}} = 325$ plane waves, corresponding to a full scan of the immersion MO pupil (NA=0.8, refractive index $n_0=1.33$). The interferograms are recorded by $N_{\textrm{out}}=512^2$ pixels of the camera, corresponding to an output FOV of $\Omega_{out} \times \Omega_{out} = 179.2 \times 179.2$ $\mu$m$^2$, with a spatial sampling $\delta \rho_\textrm{out} = 350$ nm. Finally, $N_{\omega}=180$ independent frequencies are used to probe the sample. A RPSF maximization is performed to tune the coherence plane with the focal plane in order to get the digital confocal image displayed in Fig.~\ref{fig:mire}c. A local compensation of wave distortions is performed over reduced spatial windows of size $L=9.3$ $\mu$m in order to get the final image displayed in  Fig.~\ref{fig:mire}d.}\\

\noindent\textbf{Data availability.} Optical data used in this manuscript have been deposited at Zenodo (\href{https://zenodo.org/record/8407618}{https://zenodo.org/record/8407618}).
\\

\noindent\textbf{Code availability.}
\corr{Codes used to post-process the optical data within this paper are available upon request at Zenodo (\href{https://zenodo.org/records/10674114}{https://zenodo.org/records/10674114})}\\

\noindent\textbf{Acknowledgments.}
The authors wish to thank A. Badon for initial discussions on the project, K. Irsch for providing the corneal sample, A. Le Ber for providing the iterative phase reversal algorithm and F. Bureau for his help on Supplementary movies.\\

\noindent\textbf{Funding Information.}
The authors are grateful for the funding provided by the European Research Council (ERC) under the European Union's Horizon 2020 research and innovation program (grant agreement nos. 610110 and 819261, HELMHOLTZ* and REMINISCENCE projects, respectively). This project has also received funding from Labex WIFI (Laboratory of Excellence within the French Program Investments for the Future; ANR-10-LABX-24 and ANR-10-IDEX-0001-02 PSL*) \alex{and from CNRS Innovation (Prematuration program, MATRISCOPE project)}.\\

\noindent\textbf{Author Contributions.}
A.A. initiated and supervised the project. P.B., V.B. and A.A. designed the experimental setup. P.B., V.B. and N.R. built the experimental set up. V.B. designed the acquisition scheme. V.B., P.B., N.G. and E.A. developed the post-processing tools. P.B. performed the corneal imaging experiment. V.B., P.B. and A.A. analyzed the experimental results. V.B., P.B. and A.A. performed the theoretical study. P.B., V.B. and A.A. prepared the manuscript. P.B., V.B., N.G., A.C.B., M.F. and A.A. discussed the results and contributed to finalizing the manuscript. \\

\noindent\textbf{Competing interests.}
P.B., V.B., M.F., C.B. and A.A. are named inventors on french patent FR2207334 (filing date 18.07.2022), which is related to the techniques described in this Article.

\clearpage

\renewcommand{\thetable}{S\arabic{table}}
\renewcommand{\thefigure}{S\arabic{figure}}
\renewcommand{\theequation}{S\arabic{equation}}
\renewcommand{\thesection}{S\arabic{section}}

\setcounter{equation}{0}
\setcounter{figure}{0}
\setcounter{section}{0}

\begin{center}
\huge{\bf{Supplementary Material}}
\end{center}
\normalsize
\vspace{5 mm}

This document provides further information on: (\textit{i}) the experimental set up; (\textit{ii}) 3D images of the cornea; (\textit{iii}) the theoretical expression of the multi-spectral reflection matrix;  (\textit{iv}) the theoretical expression of the focused reflection matrix; {(\textit{v}) the reflection point spread function; (\textit{vi}) the local distortion matrix; (\textit{vii}) the corresponding correlation matrix; (\textit{viii}) iterative phase reversal; (\textit{ix}) the proof-of-concept experiment with a resolution target placed behind the cornea.

\begin{figure}
    \includegraphics[width=0.8\columnwidth]{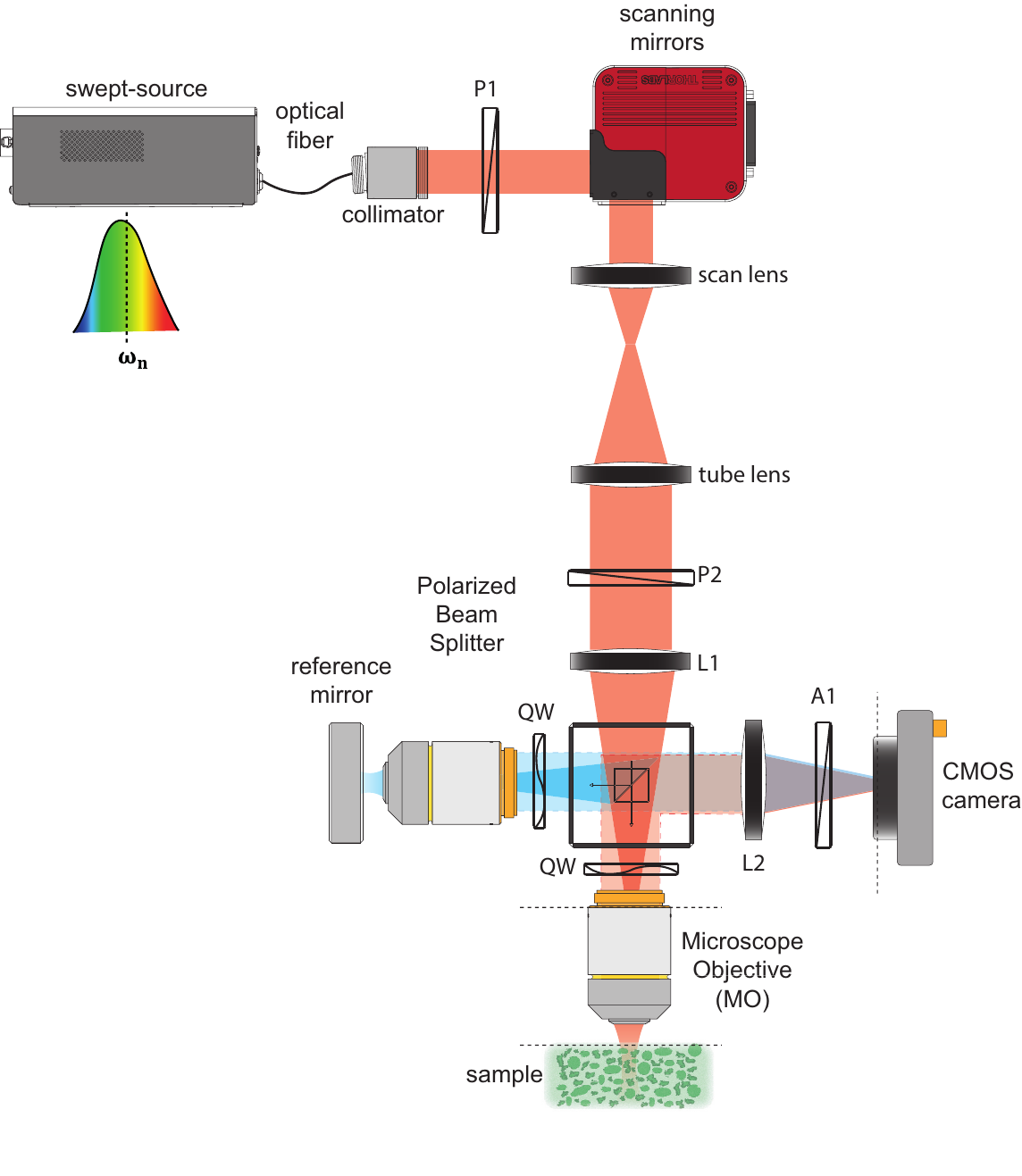}
    \caption{\textbf{Detailed experimental set up}. P: Polarizers; A: Analyzer; QW: quarter wave plates.}
    \label{fig:setup_full}
\end{figure}

\section{ Experimental set up} 

The full experimental set up is displayed in Fig.~\ref{fig:setup_full}. Compared with Fig.~1 of the accompanying paper, it shows the presence of a scan lens and of a tube lens in order to focus the incident light in the pupil plane of the microscope objectives at input. It also highlights the control of light polarization in order to minimize spurious reflections. The beam splitter is polarized and quarter wave plates are placed in the two arms such that the reflected light in the two arms is transmitted to the CMOS camera. The amount of light injected in the two arms is controlled by means of two polarizers, P1 and P2, placed before the polarized beam splitter. An analyzer A1 allows us to project the sample and reference beams on the same polarization and make them interfere in the camera plane.

\section{Other 3D images of the cornea}
In complement of Fig.~2 of the accompanying paper, Figs.~\ref{image}b$_1$-d$_1$ shows longitudinal and transverse cross-sections of the cornea obtained via spectra domain OCT (Fig.~\ref{image}a$_1$). Its comparison with the holoscopic image displayed in Figs.~2b$_1$-d$_1$ of the accompaying paper illustrate the effect of numerical focusing. While the OCT image is completely blurred due to multiple scattering and finite depth-of-field of the high-NA microsocpe objective, the numerical back-focusing of the optical wave-field improves the image contrast. Nevertheless, although the brightest are revealed, the holoscopic image still suffers from a strong multiple scattering background that generates a random speckle. This speckle can be smoothed by an incoherent average of the holoscopic image obtained for each illumination (Figs.~\ref{image}a$_2$-d$_2$). However, such an incoherent compound image remains largeley perfectible since the contrast remains very weak. On the contrary, a coherent compound of holoscopic images provides a much contrasted image of the cornea (Figs.~2a$_2$-d$_2$ of the accompanying paper). Indeed, a coherent combination of multi-illuminations acts as a confocal pinhole.

\begin{figure}
    \includegraphics[width=\textwidth]{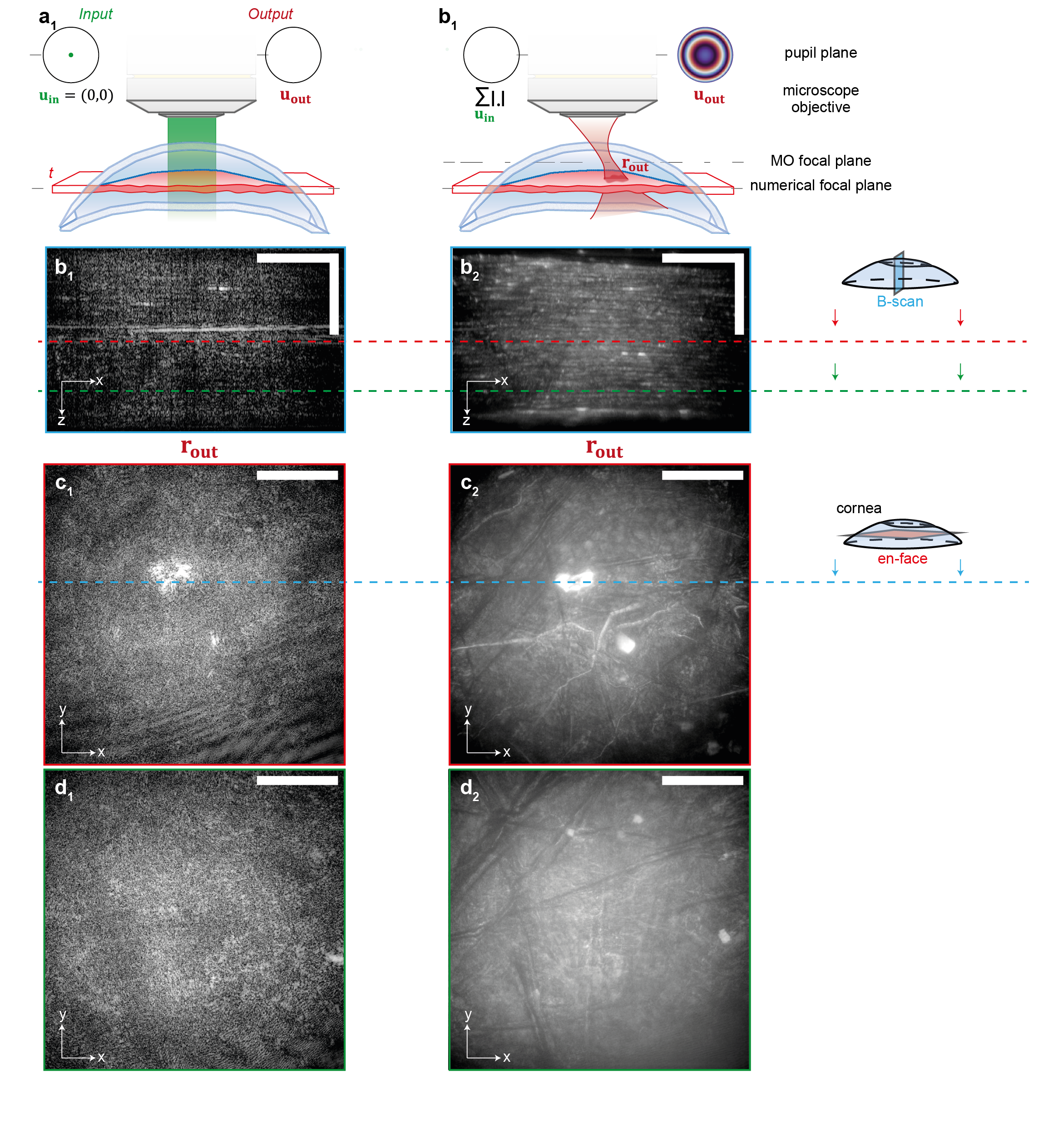}
    \caption{\textbf{Intermediate 3D images of the cornea}. \textbf{a}, Imaging Methods. \textbf{b}, B-scan image showing one longitudinal section of the cornea reflectivity.  \textbf{c}-\textbf{d}, En-face image of the cornea at $z=150$ $\mu$m and 275 $\mu$m, respectively [scale bar: 75 $\mu$m]. In panels (\textbf{a})-(\textbf{d}), subscripts 1 and 2 stand for spectral domain OCT and multi-illumination holoscopy (incoherent compound).}
    \label{image}
\end{figure}

\section{Multi-spectral reflection matrix}

In this section, we express theoretically  the multi-spectral reflection matrix recorded by the experimental set up of Fig.~\ref{fig:setup_full}. To that aim, we will rely on a simple Fourier optics model to describe the multi-spectral reflection matrix. For the sake of simplicity, this model is scalar. 

The wave field $E_s(\mathbf{s}_\textrm{out},\mathbf{u}_\textrm{in},\omega)$ reflected by the sample arm in the camera plane can be expressed as follows:
\begin{equation}
\label{Esample}
    E_s(\mathbf{s}_\textrm{out},\mathbf{u}_\textrm{in},\omega) = S(\omega)\iiint  G(\mathbf{s}_\textrm{out},\bm{\rho}_s,z_s,\omega) \gamma(\bm{\rho}_s,z_s)  E_\textrm{in}(\mathbf{u}_\textrm{in},\bm{\rho}_s,z_s,\omega)  \mathrm{d}\bm{\rho}_s dz_s.
\end{equation}
$S(\omega)$ is the amplitude of light source at frequency $\omega$. $G(\mathbf{s}_\textrm{out},\bm{\rho}_s,z_s) $, the Green's function between the sample mapped by the vector $(\bm{\rho}_s,z_s)$ and the CCD sensors identified by $\mathbf{s}_\textrm{out}$. $\gamma(\bm{\rho}_s,z_s)$ represents the sample reflectivity. $E_\textrm{in}(\mathbf{u}_\textrm{in},\bm{\rho}_s,z,\omega)$ describes the incident wave-field that can be expressed as follows: 
\begin{equation}
\label{E0}
    E_\textrm{in}(\mathbf{u}_\textrm{in},\bm{\rho}_s,z_s,\omega) = S(\omega) \mathcal{F}(\mathbf{u}_\textrm{in},z_s,\omega)  T(\mathbf{u}_\textrm{in},\bm{\rho}_s,z) \exp \left [ \alex{-} i \frac{2\pi}{\lambda f}\mathbf{u}_\textrm{in} . \bm{\rho}_s \right ] 
\end{equation}
where $\mathcal{F}(\mathbf{u}_\textrm{in},z_s,\omega)$ is the Fresnel phase law that describes plane wave propagation inside an homogeneous medium of effective index $n$ (\alex{Eq.~20} in the accompanying paper). The transmission matrix $\mathbf{T}=\left[T(\mathbf{u}_\textrm{in},\bm{\rho}_s,z) \right ]$ accounts for the wave distortions induced by the fluctuations of the optical index inside the medium. 

In the reference arm, a mirror is placed in the focal plane of the MO and displays a uniform reflectivity: $\gamma(\bm{\rho}_m,z_m)=\gamma_m \delta (z_m)$, with $\delta$ the Dirac distribution and $\gamma_m$ the mirror surface reflectivity. The reference wave-field is thus given by:
\begin{equation}
\label{Eref}
    E_{\textrm{ref}}(\mathbf{s}_\textrm{out},\mathbf{u}_\textrm{in},\omega) = \gamma_m S(\omega) \iint   G_\textrm{ref}(\mathbf{s}_\textrm{out},\bm{\rho}_m)   T_\textrm{ref}(\mathbf{u}_\textrm{in},\bm{\rho}_m)  \exp \left ( \alex{-}i \frac{2\pi}{\lambda f}\mathbf{u}_\textrm{in} . \bm{\rho}_m \right )  \mathrm{d}\bm{\rho}_m.
\end{equation}
with $G_\textrm{ref}(\mathbf{s}_\textrm{out},\bm{\rho}_m,\omega) $, the Green's functions between the focal plane of the MO ($\bm{\rho}_m$) and the CCD sensors ($\mathbf{s}_\textrm{out}$) and $T_\textrm{ref}(\mathbf{u}_\textrm{in},\bm{\rho}_m)$, the transfer function describing the aberrations undergone by the incident wave in the reference arm due to experimental imperfections (MO, misalignment, \textit{etc.}). Assuming isoplanicity in the reference arm, the Green's function $G_\textrm{ref}$ can be replaced by a spatially-invariant impulse response ${H}_\textrm{ref}$ between the focal plane and the CCD sensors: $G_{ref}(\mathbf{s}_\textrm{out},\bm{\rho}_m) = H_{ref}(\bm{\rho}_m \alex{+} \mathbf{s}_\textrm{out})$. Under the same hypothesis, the transfer function $ T_\textrm{ref}(\mathbf{u}_\textrm{in},\bm{\rho}_m)$ becomes an aberration transmittance $\mathcal{T}_\textrm{ref}(\mathbf{u}_\textrm{in})$, defined as the Fourier transform of the reference arm point spread function $ {H}_\textrm{ref}$:
\begin{equation}
   \mathcal{T}_\textrm{ref} (\mathbf{u}_\textrm{in})= \iint d\bm{\rho} {H}_\textrm{ref} (\bm{\rho}) \exp \left ( \alex{-} i \frac{2\pi}{\lambda f}\mathbf{u}_\textrm{in} \cdot \bm{\rho} \right ).
\end{equation}
Under the isoplanatic assumption, Equation~\ref{Eref} thus simplifies into:
\begin{equation}
\label{Eref2}
   E_{\textrm{ref}}(\mathbf{s}_\textrm{out},\mathbf{u}_\textrm{in},\omega) =  \gamma_m S(\omega) \mathcal{T}_\textrm{ref}(-\mathbf{u}_\textrm{in})   \mathcal{T}_\textrm{ref}(\mathbf{u}_\textrm{in})  \exp \left ( i \frac{2\pi}{\lambda f}\mathbf{u}_\textrm{in} . \mathbf{s}_\textrm{out} \right )
\end{equation}
If aberrations in the reference arm are neglected, we retrieve the fact the reference wave-field is a replica of the input wave-front:
\begin{equation}
\label{Eref3}
   E_{\textrm{ref}}(\mathbf{s}_\textrm{out},\mathbf{u}_\textrm{in},\omega) =  \gamma_m  S(\omega) \exp \left ( i \frac{2\pi}{\lambda f}\mathbf{u}_\textrm{in} . \mathbf{s}_\textrm{out} \right ).
\end{equation}
In the following, we will not make this assumption and will consider the more general expression of $E_{\textrm{ref}}$ given in Eq.~\ref{Eref2}.

The coefficients of the multi-spectral matrix $\mathbf{R}_{\mathbf{su}}(\omega)$ are recorded by isolating the interference between the sample beam, $E_s$, and the reference beam, $E_\textrm{ref}$ (Eqs.~15 and 16 of the accompanying paper):
\begin{equation}
\label{Rsu}
    R(\mathbf{s}_\textrm{out},\mathbf{u}_\textrm{in},\omega) = E(\mathbf{s}_\textrm{out},\mathbf{u}_\textrm{in},\omega) E^*_{\textrm{ref}}(\mathbf{s}_\textrm{out},\mathbf{u}_\textrm{in},\omega) \exp \left( \alex{i} \frac{2\pi}{\lambda f} \mathbf{u}_\textrm{in} . \mathbf{s}_\textrm{out}  \right ) 
\end{equation}
Using Eqs.~\ref{Esample}, \ref{E0}, \ref{Eref2}, the last equation can be rewritten as follows:
\begin{eqnarray}
\label{Rmulti}
    R(\mathbf{s}_\textrm{out},\mathbf{u}_\textrm{in},\omega)  = \gamma_m |S(\omega)|^2&\iiint&  G(\mathbf{s}_\textrm{out},\bm{\rho}_s,z_s,\omega) \gamma(\bm{\rho}_s,z_s) \nonumber \\
  &\times &  T(\mathbf{u}_\textrm{in},\bm{\rho}_s,z_s) \mathcal{F}(\mathbf{u}_\textrm{in},z_s,\omega)  \nonumber \\
  & \times &
  \mathcal{T}_\textrm{ref}^{\alex{*}}(-\mathbf{u}_\textrm{in})   \mathcal{T}_{\textrm{ref}}^{\alex{*}}(\mathbf{u}_\textrm{in})  \exp \left [\alex{-} i \frac{2\pi}{\lambda f}\mathbf{u}_\textrm{in} . \bm{\rho}_s \right ]  \mathrm{d}\bm{\rho}_s dz_s .
\end{eqnarray}
One can already notice from this expression that the aberrations undergone by the reference wave-field ($\mathcal{T}_\textrm{ref}^{\alex{*}}(-\mathbf{u}_\textrm{in})   \mathcal{T}_{\textrm{ref}}^{\alex{*}}(\mathbf{u}_\textrm{in}) $) emerge at the input of the recorded reflection matrix. 

\section{Focused reflection matrix}

In this section, we describe theoretically the numerical focusing process leading to a time-gated focused reflection matrix at each depth of the sample. 

First, a spatial Fourier transform over the camera pixels $\mathbf{s}_\textrm{out}$ leads to the reflection matrix $\mathbf{R}_{\mathbf{uu}}$ in the pupil basis \alex{(Eqs.~17 and 18 of the accompanying paper)}:
\begin{eqnarray}
\label{Rmulti2}
    R(\mathbf{u}_\textrm{out},\mathbf{u}_\textrm{in},\omega)  = \gamma_m |S(\omega)|^2 &\iiint&  T(\mathbf{u}_\textrm{out},\bm{\rho}_s,z) \mathcal{F}(\mathbf{u}_\textrm{out},z_s,\omega) \gamma(\bm{\rho}_s,z_s) \nonumber \\
  &\times &  T(\mathbf{u}_\textrm{in},\bm{\rho}_s,z) \mathcal{F}(\mathbf{u}_\textrm{in},z_s,\omega)  \nonumber \\
  & \times &
  \mathcal{T}^*_{ref}(-\mathbf{u}_\textrm{in})   \mathcal{T}^*_\textrm{ref}(\mathbf{u}_\textrm{in}) \\
  &\times  & \exp \left [\alex{-} i \frac{2\pi}{\lambda f}(\mathbf{u}_\textrm{in}+\mathbf{u}_\textrm{out}) . \bm{\rho}_s \right ]  \mathrm{d}\bm{\rho}_s dz_s.
\end{eqnarray}
As for incident light (Eq.~\ref{E0}), the return path is decomposed in the plane wave basis as the product between a Fresnel phase law $\mathcal{F}(\mathbf{u}_\textrm{out},z_s,\omega) $, accounting for free-space wave propagation in an homogeneous medium of refractive index $n$, and the transfer function $T(\mathbf{u}_\textrm{in},\bm{\rho}_s,z_s)$ that grasps the wave distortions induced by the refractive index fluctuations such that:
\begin{equation}
T(\mathbf{u}_\textrm{out},\bm{\rho}_s,z_s) \mathcal{F}(\mathbf{u}_\textrm{out},z_s,\omega) = \iint  G(\mathbf{s}_\textrm{out},\bm{\rho}_s,z_s) \exp \left ({-} i \frac{2\pi}{\lambda f}\mathbf{u}_\textrm{out}.\mathbf{s}_\textrm{out}  \right ) d \bm{\rho}_s.
\end{equation}

Numerical focusing at depth $z$ \alex{(Eqs.~1 and 4 of the accompanying paper)} then consists in compensating wave diffraction by applying the phase conjugate of the Fresnel propagator for refractive index $n_0$ at input and output before a spectral Fourier transform:
\begin{eqnarray*}
 R(\bm{\rho}_\textrm{out},\bm{\rho}_\textrm{in},z,t)=  \gamma_m \sum_{\mathbf{u}_\textrm{in}}\sum_{\mathbf{u}_\textrm{out}} & & \int d{\omega} |S({\omega}) |^2 \exp \left ( i \omega t\right ) \mathcal{F}_0^*(\mathbf{u}_\textrm{out},z,\omega) R(\mathbf{u}_\textrm{out},\mathbf{u}_\textrm{in},\omega )\mathcal{F}_0^*(\mathbf{u}_\textrm{in},z,\omega) \nonumber \\
 & &\times 
 \exp \left [ i \frac{2\pi}{\lambda f}(\mathbf{u}_\textrm{in} \cdot \bm{\rho}_\textrm{in} +\mathbf{u}_\textrm{out}\cdot \bm{\rho}_\textrm{out}) . \bm{\rho}_s \right ] 
\end{eqnarray*}
Injecting Eq.~\ref{Rmulti2} leads to the following expression for the coefficients of $\mathbf{R}_{\bm{\rho\rho}}(z,\omega)$:
\begin{align}
\label{Rmulti3}
     R(\bm{\rho}_\textrm{out},\bm{\rho}_\textrm{in},z,t)  = & \gamma_m \sum_{\mathbf{u}_\textrm{in}} \sum_{\mathbf{u}_\textrm{out}}  \int d{\omega} |S({\omega}) |^2 \exp \left ( i \omega t\right )  \nonumber\\ 
		& \times \iiint 
    T(\mathbf{u}_\textrm{out},\bm{\rho}_s,z)  \gamma(\bm{\rho}_s,z_s,\omega)   T(\mathbf{u}_\textrm{in},\bm{\rho}_s,z_s) \mathcal{T}^*_\textrm{ref}(-\mathbf{u}_\textrm{in})   \mathcal{T}^*_\textrm{ref}(\mathbf{u}_\textrm{in})\nonumber \\
  &\times  \mathcal{F}(\mathbf{u}_\textrm{out},z_s,\omega) \mathcal{F}_0^*(\mathbf{u}_\textrm{out},z,\omega)   \mathcal{F}(\mathbf{u}_\textrm{in},z_s,\omega) \mathcal{F}_0^*(\mathbf{u}_\textrm{in},z,\omega) \nonumber \\  
  &\times  \exp \left \lbrace - i \frac{2\pi}{\lambda f} \left [\mathbf{u}_\textrm{in} . (\bm{\rho}_s-\bm{\rho}_\textrm{in}) + \mathbf{u}_\textrm{out} . (\bm{\rho}_s-\bm{\rho}_\textrm{out}) \right ] \right \rbrace
  \mathrm{d}\bm{\rho}_s dz_s.
  \end{align}
The positions of the coherence volume and focusing plane are determined by the cancellation of the Fresnel phase laws, 
\begin{equation}
\label{fresnel}
\begin{split}
& \mathcal{F}(\mathbf{u}_\textrm{in},z,\omega) \mathcal{F}_0^*(\mathbf{u}_\textrm{in},z_s,\omega) \mathcal{F}(\mathbf{u}_\textrm{out},z,\omega) \mathcal{F}_0^*(\mathbf{u}_\textrm{out},z_s,\omega) = \\
& \exp \left [ i \frac{\omega}{c_0} \left ( \sqrt{n_0^2 - \frac{||\mathbf{u}_\textrm{in} ||^2}{f^2}} + \sqrt{n_0^2 - \frac{||\mathbf{u}_\textrm{out} ||^2}{f^2}} - 2n_0 \right ) z \right ]\\
& \times \exp \left [ - i \frac{\omega}{c_0} \left ( \sqrt{n^2 - \frac{||\mathbf{u}_\textrm{in} ||^2}{f^2}} + \sqrt{n^2 - \frac{||\mathbf{u}_\textrm{out} ||^2}{f^2}}   \right ) z_s\right ].
\end{split}
\end{equation}
Under the paraxial approximation, these Fresnel phase laws can be developed as follows: 
\begin{equation}
\label{fresnel2_supp}
\begin{split}
& \mathcal{F}(\mathbf{u}_\textrm{in},z,\omega) \mathcal{F}_0^*(\mathbf{u}_\textrm{in},z_s,\omega) \mathcal{F}(\mathbf{u}_\textrm{out},z,\omega) \mathcal{F}_0^*(\mathbf{u}_\textrm{out},z_s,\omega) = \\
& \exp \left [ -2i \frac{n\omega}{c_0}  z_s  \right ] \exp \left [ -i \frac{\omega}{c_0} \left ( \frac{||\mathbf{u}_\textrm{in} ||^2}{2f^2} +  \frac{||\mathbf{u}_\textrm{out} ||^2}{2f^2}\right) \left (\frac{z}{n_0}-\frac{z_s}{n} \right ) \right ],
\end{split}
\end{equation}
The cancellation of the first phase term defines the real position of the coherence volume $z_s=z_t=c_0t/(2n)$ that appears at an effective depth $z_0=c_0t/(2n_0)=(n/n_0)z_t$ (Fig.~\ref{fig:defocus}a). Previous expression of $R(\bm{\rho}_\textrm{out},\bm{\rho}_\textrm{in},z,t) $ (Eq.~\ref{Rmulti3}) can be rewritten as follows:
\begin{align}
\label{Rmulti4}
     R(\bm{\rho}_\textrm{out},\bm{\rho}_\textrm{in},z,t)  = & \gamma_m f(t-2 n z_t/c) \stackrel{t}{\circledast}\\
     & \sum_{\omega}\sum_{\mathbf{u}_\textrm{in}} \sum_{\mathbf{u}_\textrm{out}} \iiint 
    T(\mathbf{u}_\textrm{out},\bm{\rho}_s,z)  \gamma(\bm{\rho}_s,z_s) \\
    &\times  T(\mathbf{u}_\textrm{in},\bm{\rho}_s,z_s) \mathcal{T}^*_\textrm{ref}(-\mathbf{u}_\textrm{in})   \mathcal{T}^*_\textrm{ref}(\mathbf{u}_\textrm{in}) \exp \left ( i \omega t\right )\nonumber \\
  &\times   \exp \left [ -i \frac{\omega}{c_0} \left ( \frac{||\mathbf{u}_\textrm{in} ||^2}{2f^2} +  \frac{||\mathbf{u}_\textrm{out} ||^2}{2f^2}\right) \left (\frac{z}{n_0}-\frac{z_s}{n} \right ) \right ] \nonumber \\  
  &\times  \exp \left \lbrace - i \frac{2\pi}{\lambda f} \left [\mathbf{u}_\textrm{in} . (\bm{\rho}_s-\bm{\rho}_\textrm{in}) + \mathbf{u}_\textrm{out} . (\bm{\rho}_s-\bm{\rho}_\textrm{out}) \right ] \right \rbrace
  \mathrm{d}\bm{\rho}_s dz_s.
  \end{align}
with $f(t)=\int d\omega |S(\omega)|^2 e^{i \omega t}$, the time response of the microscope. The symbol $\circledast$ stands for convolution over variable $t$. The position $z_f$ of the focusing plane is obtained when the parabolic phase term cancels in the previous expression, that is to say for $z_s=z_f=(n_0/n)z_t$. The apparent defocus induced by the mismatch between $n$ and $n_0$ is thus equal to:
\begin{equation}
\Delta z=z_f-z_0 = z_t (n_0/n - n/n_0) 
\end{equation}
An index mismatch thus implies a defocus distance $\Delta z$ that increases linearly with $z_t$ (Fig.~\ref{fig:defocus}a). In the present study, the estimated defocus is roughly constant with $z_t$. It thus means that the cornea displays an effective optical index $n\sim n_0 =1.33$ and that the observed defocus rather originates from the different lengths between the sample and reference arms.

Once the focusing plane is matched with the coherence volume, the Fresnel phase laws in Eq.~\ref{Rmulti4} vanish. Assuming $n=n_0$, the coefficients of the time-gated focused reflection matrix $\mathbf{R}_{\bm{\rho \rho}}(z_t)$ can be derived as follows:
\begin{align}
\label{Rrr0}
     R(\bm{\rho}_\textrm{out},\bm{\rho}_\textrm{in},z_t)  = & \gamma_m \sum_{\mathbf{u}_\textrm{in}} \sum_{\mathbf{u}_\textrm{out}} \iint 
    T_\textrm{out}(\mathbf{u}_\textrm{out},\bm{\rho}_s,z_t)  \gamma(\bm{\rho}_s,z_t) T_\textrm{in}(\mathbf{u}_\textrm{in},\bm{\rho}_s,z_t) \nonumber \\
  &\times  \exp \left \lbrace i \frac{2\pi}{\lambda f} \left [\mathbf{u}_\textrm{in} . (\bm{\rho}_s-\bm{\rho}_\textrm{in}) + \mathbf{u}_\textrm{out} . (\bm{\rho}_s-\bm{\rho}_\textrm{out}) \right ] \right \rbrace
  \mathrm{d}\bm{\rho}_s .
  \end{align}
 with 
 $$ T_\textrm{out}(\mathbf{u}_\textrm{out},\bm{\rho}_s,z_t) \equiv T(\mathbf{u}_\textrm{out},\bm{\rho}_s,z_t)$$ and $$ T_\textrm{in}(\mathbf{u}_\textrm{in},\bm{\rho}_s,z_t) \equiv T(\mathbf{u}_\textrm{in},\bm{\rho}_s,z_t) \mathcal{T}^*_{ref}(-\mathbf{u}_\textrm{in})   \mathcal{T}^*_\textrm{ref}(\mathbf{u}_\textrm{in}).$$ While the output transmission matrix $\mathbf{T}_\textrm{out}$ corresponds to the sample transmission matrix $\mathbf{T}$, the input transmission matrix $\mathbf{T}_\textrm{in}$ grasps both the sample and reference arm aberrations. 

\begin{figure}
    \includegraphics[width=0.6\textwidth]{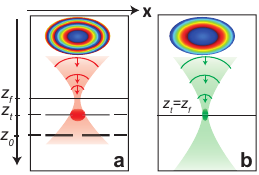}
    \caption{\textbf{Mismatch between the coherence volume and focusing plane}. \textbf{a}. For a medium of refractive index $n>n_0$, the focusing plane at $z_f$ is shifted from the coherence volume at $z_t$ and expected ballistic depth $z_0$. \textbf{b}. A defocus $\Delta z$ can be applied in post-processing in order to make coincide the coherence and focusing planes.}
    \label{fig:defocus}
\end{figure}

 The Fourier transform of the transmission coefficients $T_\textrm{in/out}$ in Eq.~\ref{Rrr0} provide local PSFs, $H_\textrm{in/out}$, such that:
 \begin{equation}
\label{HH}
H_\textrm{in/out}(\bm{\rho},\bm{\rho}_s,z_t)= \sum_{\mathbf{u}}T_\textrm{in/out} (\mathbf{u},\bm{\rho_s},z_t) \exp \left ( i \frac{2\pi}{\lambda f} \mathbf{u} . \bm{\rho}  \right ) .
 \end{equation}
The time-gated reflection matrix can be rewritten as follows:
 \begin{equation}
\label{Rrr}
     R(\bm{\rho}_\textrm{out},\bm{\rho}_\textrm{in},z_t)  = \iint d \bm{\rho}_s
    H_\textrm{out}(\bm{\rho}_s-\bm{\rho}_\textrm{out},\bm{\rho}_s,z_t)  \gamma(\bm{\rho}_s,z) H_\textrm{in}(\bm{\rho}_s-\bm{\rho}_\textrm{in},\bm{\rho}_s,z_t) 
  \end{equation}
The latter expression can be recast as a function the impulse responses $G_\textrm{in/out}(\bm{\rho}_s,\bm{\rho}_\textrm{in/out},z_t)$ between input/output focusing points and points $\bm{\rho}_s$ mapping the sample. Both quantities are actually linked as follows: 
\begin{equation}
G_\textrm{in/out}(\bm{\rho}_s,\bm{\rho}_\textrm{in/out},z_t)=H_\textrm{in/out}(\bm{\rho}_s-\bm{\rho}_\textrm{in/out},\bm{\rho}_s,z_t)
\end{equation}
Injecting the last expression into Eq.~\ref{Rrr} leads to the following expression:
 \begin{equation}
\label{Rrr2}
     R(\bm{\rho}_\textrm{out},\bm{\rho}_\textrm{in},z_t)  = \iint 
    G_\textrm{out}(\bm{\rho}_\textrm{out},\bm{\rho}_s,z_t)  \gamma(\bm{\rho}_s,z_t) G_\textrm{in}(\bm{\rho}_\textrm{in},\bm{\rho}_s,z_t) 
  \end{equation}
Under a matrix formalism, the last expression can be rewritten as follows:
\begin{equation}
\mathbf{R}_{\bm{\rho \rho}}(z_t)=\mathbf{G}_\textrm{out} (z_t)\times \bm{\Gamma}(z_t) \times  \mathbf{G}^\top_\textrm{in}(z_t),
\end{equation}
$\bm{\Gamma}$ describes the scattering process inside the sample. Under a single scattering assumption, this matrix is diagonal. Its coefficients then correspond the sample reflectivity $\gamma(\bm{\rho}_s,z_t)$. $\mathbf{G}_\textrm{in}$ and $\mathbf{G}_\textrm{out}$ are the input and output focusing matrices. Their coefficients, $G_\textrm{in/out}(\bm{\rho}_\textrm{in/out},\bm{\rho}_s,z_t) $, describe the transverse amplitude distribution of the focal spot when trying to focus at point $(\bm{\rho}_\textrm{in/out},z_t)$.

\section{Reflection point spread function}

As mentioned in the accompanying paper, the off-diagonal cofficients of $\mathbf{R}_{\bm{\rho \rho}}(z,t)$ enable to probe the focusing quality at any voxel by investigating the reflection point spread function (RPSF, \alex{Eq.~6}). To express theoretically the latter quantity, a local isoplanatic assumption shall be made. This hypothesis implies that the PSFs ${H}_\textrm{in/out}$ are locally invariant by translation. This leads us to define local spatially-invariant PSFs ${h_\textrm{in/out}}$ around each central midpoint $\bm{\rho}_p$ at each time-of-flight such that: 
\begin{equation}
\label{local_iso}
H_\textrm{in/out}(\bm{\rho}_s-\bm{\rho}_\textrm{in/out},\bm{\rho}_s,z,t) =h_\textrm{in/out}(\bm{\rho}_s-\bm{\rho}_\textrm{in/out},\bm{\rho}_p,z_p,t).
\end{equation}
The second assumption is to consider the medium reflectivity $\gamma(\bm{\rho}_s,z)$ as random:
\begin{equation}
\label{gamma_random}
\left\langle\gamma\left(\bm{\rho}_1{,z}\right) \gamma^*\left(\bm{\rho}_2{,z}\right)\right\rangle=\left\langle|\gamma|^2\right\rangle \delta\left(\bm{\rho}_2-\bm{\rho}_1\right),
\end{equation}
By combining those assumptions with Eq.~\ref{Rrr}, the ensemble average of $RPSF(\Delta {\bm{\rho}},\bm{\rho}_\textrm{p}, z,t)$ \alex{(Eq.~7)} can be expressed as follows:
\begin{equation}
\left \langle RPSF(\Delta {\bm{\rho}}, \bm{\rho}_\textrm{p}, z,t) \right \rangle =\left\langle|\gamma|^2\right\rangle  \times \left[\left|h_{\textrm{in}}\right|^2 \stackrel{\Delta {\bm{\rho}}}{\circledast}\left|h_{\textrm{out}}\right|^2\right](\Delta {\bm{\rho}},\bm{\rho}_\textrm{p}, z_p,t) .
\end{equation}

\section{Local distortion matrix}

Wave distortions can be investigated both at input and output of the reflection matrix (see Methods of the accompanying paper). Here we will consider the properties of the output distortion matrix but the same theoretical developments can be made at input.

The output distortion matrix can be built by first projecting the time-gated focused reflection matrix $\mathbf{R}_{\bm{\rho \rho}}(z_t)$ in the pupil plane at output:
\begin{equation}
    \mathbf{R}_{\mathbf{u}\bm{\rho}}(z_t) = \mathbf{P}_{\mathbf{u}\bm{\rho}} \times \mathbf{R}_{\bm{\rho \rho}}(z_t).
\end{equation} 
Then, the distorted component of the wave-field can be extracted by subtracting the geometric phase expected in an ideal case (without aberrations). Mathematically, this can be performed using the following matrix element wise product:
\begin{equation}
    \mathbf{D}_{\textrm{out}}(z_t) =  \mathbf{R}_{\mathbf{u}\bm{\rho}}(z_t)\circ  \mathbf{P}^*_{\mathbf{u}\bm{\rho}} .
\end{equation}
or in terms of matrix coefficients,
\begin{equation}
    D(\mathbf{u}_\textrm{out},\bm{\rho}_{\textrm{in}},z_t) =\sum_{\bm{\rho}_\textrm{out}} R(\bm{\rho}_\textrm{out}, \bm{\rho}_\textrm{in},z_t)  \exp \left [ -i \frac{2\pi}{\lambda f} \mathbf{u}_\textrm{out} \cdot (\bm{\rho}_\textrm{out} - \bm{\rho}_\textrm{in}) \right ].
\end{equation}
Injecting Eqs.~\ref{HH} and \ref{Rrr} into the last equation yields
\begin{align}
\label{D}
    D(\mathbf{u}_\textrm{out},\bm{\rho}_{\textrm{in}},z_t) = \iint &T_\textrm{out}(\mathbf{u}_\textrm{out},\bm{\rho}_s,z_t) \gamma(\bm{\rho}_s,z_t) H_\textrm{in} (\bm{\rho}_s-\bm{\rho}_\textrm{in}, \bm{\rho}_s,z) \\
		& \times  \exp \left [ -i \frac{2\pi}{\lambda f} \mathbf{u}_\textrm{out} . (\bm{\rho}_s-\mathbf{\rho}_\textrm{in})\right ]  
 \textrm{d}\bm{\rho}_s .
\end{align}
In previous papers~\cite{badon_distortion_2020,Najar2023}, we showed that the distortion matrix $\mathbf{D}$ highlights spatial correlations of the reflected wave-field induced by the shift-shift memory effect~\cite{judkewitz_translation_2015,Osnabrugge2017}: Waves produced by nearby points inside \alex{an anisotropic scattering} medium generate highly correlated distorted wave-fields in the pupil plane. A strong similarity can be observed between distorted wave-fronts associated with neighboring points but this correlation tends to vanish when the two points are too far away.

To extract and exploit {this} local memory effect for imaging, the field-of-illumination should be subdivided into overlapping regions~\cite{Najar2023} that are defined by their central midpoint $(\bm{\rho}_p,z_t)$ and their lateral extension $L$. All of the distorted components associated with focusing points $\bm{\rho}_\textrm{in}$ located within each region are extracted and stored in a local distortion matrix $\mathbf{D}'_\textrm{out}(\bm{\rho}_p,z_t)$:
\begin{equation}
\label{eq:window2}
    D'(\mathbf{u}_\textrm{out},\bm{\rho}_\textrm{in}, \bm{\rho}_p,z_t) = D(\mathbf{u}_\textrm{out},\bm{\rho}_\textrm{in},z_t) ~ W_L(\bm{\rho}_\textrm{in} - \bm{\rho}_p),
\end{equation}
where $W_{L}(x,y) = 1$ for $|x|<L/2$ and $|y|<L/2$, and zero otherwise.

Under a local isoplanatic assumption (Eq.~\ref{local_iso}), the aberrations can be modelled by a local transmittance $\mathcal{T}_\textrm{out}(\mathbf{u}_\textrm{out},\bm{\rho}_\textrm{p},z_t)$ around each point $(\bm{\rho}_\textrm{p},z_t)$, such that $T_\textrm{out}(\mathbf{u}_\textrm{out},\bm{\rho}_s,z_t)\simeq \mathcal{T}_\textrm{out}(\mathbf{u}_\textrm{out},\bm{\rho}_\textrm{p},z_t)$. This transmittance is 
the Fourier transform of the local PSF ${h_\textrm{out}}(\bm{\rho},\bm{\rho}_\textrm{p},z_t)$: 
\begin{equation}
  \mathcal{T}_\textrm{out}(\mathbf{u}_\textrm{out},\bm{\rho}_\textrm{p},z_t)=\iint {h}_\textrm{out}(\bm{\rho},\bm{\rho}_\textrm{p},z) \exp \left ( \frac{2\pi}{\lambda f} \mathbf{u}_\textrm{out} \cdot \bm{\rho} \right )  \textrm{d} {\bm{\rho}} 
\end{equation}
Under this assumption, Eq.\ref{D} can be rewritten as follows:
\begin{align}
{D'}(\mathbf{u}_\textrm{out},\bm{\rho}_\textrm{in}, \bm{\rho}_p,z_t) = &
  \underbrace{\mathcal{T}_\textrm{out}(\mathbf{u}_\textrm{out},\bm{\rho}_\textrm{p},z_t)}_{\mbox{transmittance}} \label{Dalex}\\
	&\times \underbrace{\iint \gamma(\bm{\rho}+\bm{\rho}_\textrm{out},z_t) h_\textrm{in}(\bm{\rho},\bm{\rho}_\textrm{p},z_t)  \exp \left ( - i \frac{2\pi}{\lambda f}  \mathbf{u}_\textrm{in} \cdot \bm{\rho} \right ) \textrm{d} \bm{\rho}}_{\mbox{virtual source}}.
\nonumber
\end{align}
The physical meaning of this last equation is the following: Each distorted wave-field corresponds to the diffraction of a virtual source synthesized inside the medium modulated by the transmittance $\mathcal{T}_\textrm{out}$ of the sample between the focal and pupil planes. Each virtual source is spatially incoherent due to the random reflectivity of the medium, and its size is governed by the spatial extension of the input focal spot. The idea is now to smartly combine each virtual source to generate a coherent guide star and estimate $\mathcal{T}_\textrm{out}$ independently from the sample reflectivity.  

\section{Correlation of wave distortions}

To do so, the correlation matrix $\mathbf{C}_\textrm{out}=\mathbf{D}_\textrm{out}\mathbf{D}_\textrm{out}^\dag$ is an excellent tool. Its coefficients write as follows 
\begin{equation}
\label{corr_out_supp}
    C_\textrm{out} (\mathbf{u}_\textrm{out}, \mathbf{u}'_\textrm{out},\bm{\rho}_p,z_t)= N_\mathcal{W}^{-1} \sum_{\bm{\rho}_\textrm{in}} {D}'(\mathbf{u}_\textrm{out},\bm{\rho}_\textrm{in},\bm{\rho}_p,z_t){D}'^*(\mathbf{u}'_\textrm{out},\bm{\rho}_\textrm{in},\bm{\rho}_p,z_t)
\end{equation}
The matrix $\mathbf{C}_\textrm{out}(\bm{\rho}_p,z_t)$ can be decomposed as the sum of its ensemble average, the covariance matrix $\left \langle \mathbf{C}_\textrm{out}\right \rangle (\bm{\rho}_p,z_t)$, and a perturbation term $\delta \mathbf{C}_\textrm{out}(\bm{\rho}_p,z_t) $:
\begin{equation}
\label{C}
   \mathbf{C}_\textrm{out}(\bm{\rho}_p,z_t)= \left \langle \mathbf{C}_\textrm{out}\right \rangle (\bm{\rho}_p,z_t) +  \delta \mathbf{C}_\textrm{out}(\bm{\rho}_p,z_t).
\end{equation}
The intensity of the perturbation term scales as the inverse of the number $N_\textrm{L}=(L/\delta \rho_0)^2$ of resolution cells in each sub-region~\cite{robert_greens_2008,lambert_ultrasound_2022}:
\begin{equation}
\label{perturbation}
    \left \langle \left |\delta C_\textrm{out}(\mathbf{u},\mathbf{u}',\bm{\rho}_p,z_t)\right |^2 \right \rangle = \frac{ \left \langle \left | C_\textrm{out}(\mathbf{u},\mathbf{u},\bm{\rho}_p,z_t)\right |^2 \right \rangle}{N_L}
\end{equation}
This perturbation term can thus be reduced by increasing the size $L$ of the spatial window $W_L$, but at the cost of a resolution loss.

Under assumptions of local isoplanicity (Eqs.~\ref{local_iso} and \ref{Dalex}) and random reflectivity, 
\begin{equation}
\label{random}
\langle \gamma(\bm{\rho}_s,z)\gamma^*(\bm{\rho}'_s,z) \rangle= \langle | \gamma |^2 \rangle \delta (\bm{\rho}_s - \bm{\rho}'_s).
\end{equation}
The coefficients of the covariance matrix can be expressed as follows~\cite{lambert_distortion_2020}:
\begin{equation}
\label{rhoaveeq}
\left \langle \mathbf{C}_\textrm{out} \right \rangle (\bm{\rho}_p,z_t) =  \left [\mathbf{\mathcal{T}}_\textrm{out} (\bm{\rho}_p,z_t) \circ \mathbf{P}_{\mathbf{u}\bm{\rho}} \right ] \times  \mathbf{C}_H (\rp) \times \left [\mathbf{P}_{\mathbf{u}\bm{\rho}} \circ \mathbf{\mathcal{T}}_\textrm{out}(\bm{\rho}_p,z_t) \right ]^{\dag} ,
\end{equation}
or in terms of matrix coefficients,
\begin{align}
\label{rhoaveeq2}
\left \langle \mathbf{C}_\textrm{out} \right \rangle (\mathbf{u}_\textrm{out},\mathbf{u}'_\textrm{out},\bm{\rho}_p,z_t) = & \mathcal{T}_\textrm{out} (\mathbf{u}_\textrm{out},\bm{\rho}_p,z_t)  \mathcal{T}_\textrm{out}^* (\mathbf{u}'_\textrm{out},\bm{\rho}_p,z_t) \nonumber \\
 & \times  \iint \textrm{d}\bm{\rho}_s \left |h_\textrm{in}(\bm{\rho}_s,\bm{\rho}_p,z_t) \right |^2 \exp \left [-i\frac{2\pi}{\lambda f} (\mathbf{u}_\textrm{out}-\mathbf{u}'_\textrm{out})\cdot\bm{\rho}_s \right ] \nonumber \\ 
= & \mathcal{T}_\textrm{out} (\mathbf{u}_\textrm{out},\bm{\rho}_p,z_t)  \mathcal{T}_\textrm{out}^* (\mathbf{u}'_\textrm{out},\bm{\rho}_p,z_t) \nonumber \\
& \times \underbrace{ \left [\mathcal{T}_\textrm{in} \stackrel{\mathbf{u}}{\ast} \mathcal{T}_\textrm{in} \right ] (\mathbf{u}_\textrm{out}-\mathbf{u}'_\textrm{out},\bm{\rho}_p,z_t)}_{=C_H(\mathbf{u}_\textrm{out},\mathbf{u}'_\textrm{out},\bm{\rho}_p,z_t)},
\end{align}
where the symbol $\ast$ stands for correlation product over variable $\mathbf{u}$. $\mathbf{C}_H$ is a reference correlation matrix that would be measured in an homogeneous cornea for a virtual reflector whose scattering distribution corresponds to the output focal spot intensity $|h_\textrm{in}(\bm{\rho}_s,\bm{\rho}_p,z_t)|^2$.
The covariance matrix $\left \langle \mathbf{C}_\textrm{in} \right \rangle (\bm{\rho}_p,z_t)$ thus corresponds to the same experimental situation but for a virtual reflector embedded into the heterogeneous cornea under study.

\section{Iterative phase reversal}

An estimator $\exp [i{\Phi_\textrm{out}(\bm{\rho}_p,z_t)}]$ of the local aberration transmittance $\mathbf{\mathcal{T}}_\textrm{out}(\bm{\rho}_p,z_t)$ can be extracted by applying an iterative phase reversal algorithm to $\mathbf{C}_\textrm{out}$ (Eq.~34 of the accompanying paper). It mimics an iterative time reversal process on the virtual reflector described above but imposes a constant amplitude for the time-reversal invariant. This iterative phase reversal (IPR) process converges towards a wavefront that maximizes the coherence of the wave-field reflected by the virtual reflector~\cite{Najar2023}. 

This IPR process assumes the convergence of the correlation matrix $\mathbf{C}_\textrm{out}$ \alex{(Eq.~\ref{C})} towards its ensemble average $\left \langle \mathbf{C}_\textrm{out} \right \rangle $, the covariance matrix~\cite{lambert_distortion_2020,lambert_ultrasound_2022}. In fact, this convergence is never fully realized and $\mathbf{C}_\textrm{out}$ should be decomposed as the sum of this covariance matrix $\left \langle \mathbf{C}_\textrm{out}\right \rangle $ and the perturbation term $\delta \mathbf{C}_\textrm{out} $ (Eq.~\ref{C}). This incomplete convergence towards the covariance matrix leads to a phase error $\delta \phi$ made on our estimation of each aberration phase law. The variance of this error scales as follows:
\begin{equation}
\left \langle \left |\delta \phi_\textrm{out} (\mathbf{u}) \right|^2 \right \rangle \sim \frac{1}{N_L  \mathcal{C}_\textrm{in}} 
\end{equation}
 with $\mathcal{C}_\textrm{in}$, the coherence factor that is a direct indicator of the focusing quality~\cite{mallart_adaptive_1994}, ranging from 0 for a fully blurred guide star to $2/9$ for a diffraction-limited focal spot~\cite{Silverstein2001}. On the one hand, this scaling of the phase error with $N_W$ explains why each spatial window should be large enough to encompass a sufficient number of independent realizations of disorder. On the other hand, it should be small enough to grasp the spatial variations of aberration phasr laws across the field-of-view. A compromise has thus to be found between these two opposite requirements. It has led us to to choose spatial windows of size $L=18.6$ $\mu$m to ensure the convergence of the IPR process. Note that, contrary to a recent study performed in an extremely opaque cornea~\cite{Najar2023}, a multi-scale approach of wave distortions is not necessary here since the cornea under study displays smoother fluctuations of optical index over larger characteristic length scales. As previous works dealing with the CLASS algorithm~\cite{kang_high-resolution_2017,yoon_laser_2020}, these imaging conditions ensure:
 \begin{itemize}
 \item A lower level of transverse aberrations, hence a higher coherence factor that allows a direct convergence of IPR over reduced spatial windows.
 \item Larger isoplanatic patches, hence a direct convergence of the IPR process for reduced spatial windows and no need to iterate the IPR process over smaller spatial windows.
 \end{itemize}

\section{Imaging a resolution target through an opaque region of the cornea}

In addition to the three-dimensional imaging experiment of the cornea presented in the accompanying manuscript, an academic experiment has been performed and consists in imaging a resolution target behind the same cornea. This experiment provides a validation of the whole RMI process with a ground-truth object. It also allows us to test RMI in a stronger scattering regime since we image the resolution target through a more opaque region of the cornea, closer to the iris. This way, we demonstrate the ability of RMI in overcoming high-order aberrations and forward multiple scattering.

\begin{figure}[h]
    	\centering
    \includegraphics[width=\textwidth]{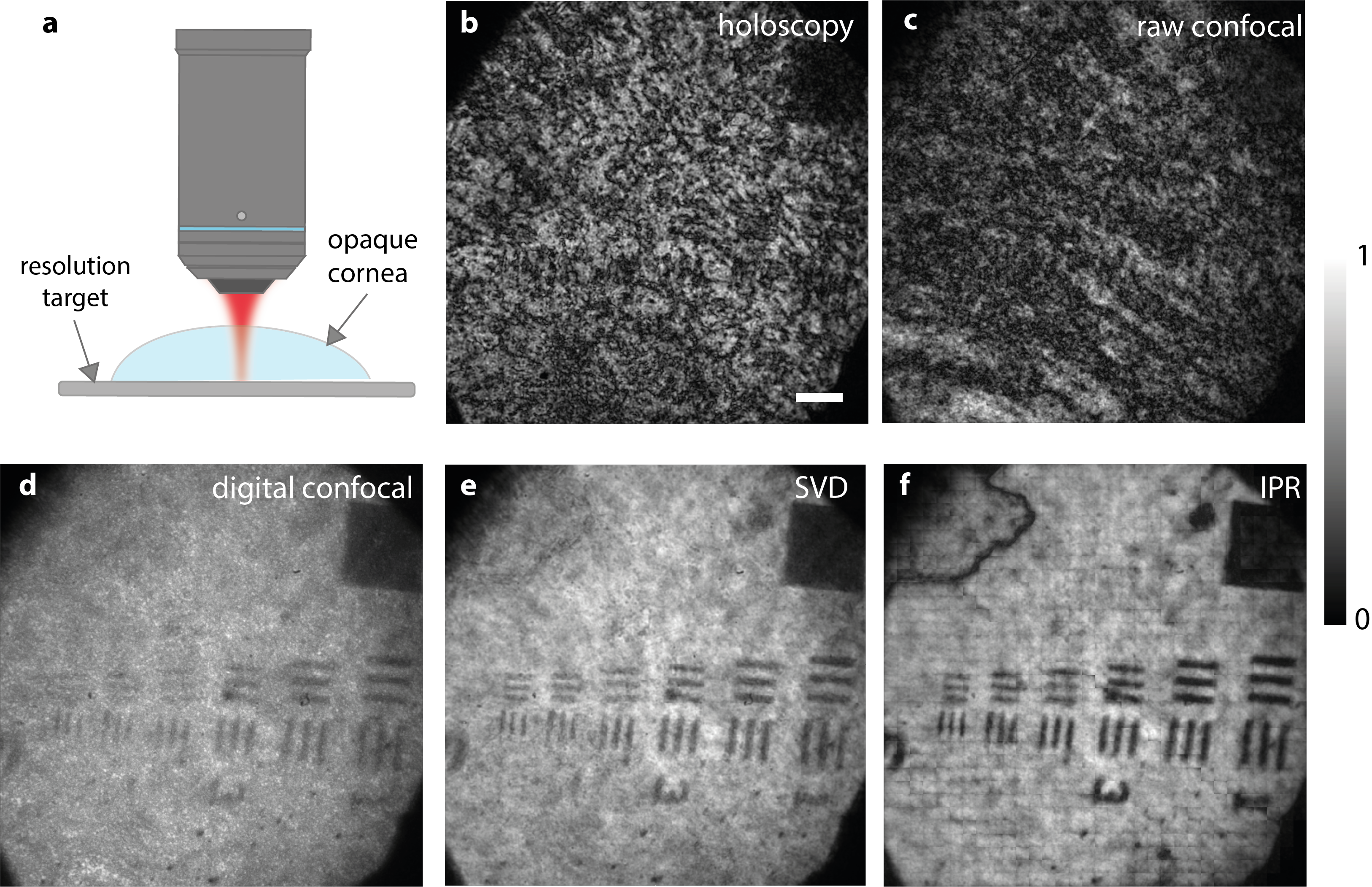}
    \caption{\textbf{Imaging a resolution target through the cornea.} \textbf{a}, Experimental configuration. \textbf{b}, Holoscopy [scale bar: 20$\mu$m]. \textbf{c}, Digital confocal image. \textbf{d}, Matrix image based on a SVD approach~\cite{badon_distortion_2020}. \textbf{e}, Matrix image based on the IPR algorithm. }
    \label{fig:image_mire}
\end{figure}

Figure~\ref{fig:image_mire}a shows the experimental configuration in which a resolution target is placed behind the cornea in the focal plane of the microscope objective (Methods). Figure~S3b displays the image obtained for a single illumination ($\mathbf{u}_\textrm{in}=\mathbf{0}$, Eq.~3). This image exhibits a random speckle due to the multiple scattering events induced by the heterogeneities of the cornea. A direct combination of each backscattered wave-field for each plane wave illumination yields a raw confocal image (Fig.~\ref{fig:image_mire}c, Eq.~5). \rev{Despite spatial multiplexing and time-gating,} this image does not provide a better result because of the mismatch between the coherence volume and the focal plane (Fig.~\ref{fig:defocus}a). An optimization of the RPSF allows us to finely tune these two planes (Fig.~\ref{fig:defocus}b)\rev{, thereby providing the digital confocal image displayed in Fig.~\ref{fig:image_mire}d.} This image reveals most of the patterns of the resolution target. However, it is poorly contrasted ($\chi\sim 0.9$ dB) because of the aberrations induced by the fluctuations of the refractive index inside the cornea. The corresponding RPSFs display a blurred feature characteristic of multiple scattering, which highlights the cornea turbidity \rev{(Fig.~5g)}. The associated Strehl ratio is extremely low: $\mathcal{S}\sim 10^{-3}$

Figure~\ref{fig:image_mire}e shows a corrected image based on a SVD decomposition of the full-field $\mathbf{D}$-matrix as proposed in a previous paper~\cite{badon_distortion_2020}. This image is built from the 25th first eigenstates of $\mathbf{D}$. It displays a better contrast than the optimized confocal image (Fig.~\ref{fig:image_mire}d) but it remains far from being perfect due to the high number of isoplanatic patches in the FOV. Note that considering more eigenstates would not improve the image contrast because the higher-rank eigenstates are polluted by multiple scattering noise. 

To circumvent this problem, an analysis of wave distortions can be performed over reduced spatial windows (Methods). The corresponding aberration phase laws are displayed in \rev{Fig.~5e,f}. They exhibit a complex feature characteristic of forward multiple scattering with a broad spatial frequency content and a short-scale memory effect~\cite{Najar2023}. The phase conjugate of the associated transmission matrix (Eq.~8) provides the final image (Eq.~9) displayed in Fig.~\ref{fig:image_mire}f. The high contrast of the image demonstrates the benefit of a local analysis of wave distorsions compared with a global SVD approach (Fig.~\ref{fig:image_mire}e). The comparison between original and final RPSFs confirms the drastic improvement of the focusing quality \rev{(Figs.~5g and h)}. The Strehl ratio is increased by a factor 14. Nevertheless, the final RPSFs are not perfect: (\textit{i}) The contrast remains limited ($\chi \sim 10$); (\textit{ii}) The transverse resolution is slighly larger than the confocal resolution $\delta \rho_c$: {$\delta \rho \sim 0.35$ $\mu$m}. Aberrations associated with an isoplanatic length smaller than $L$ cannot be addressed by RMI and could explain, at least partially, the residual flaws of the final RPSFs. Moreover, the incomplete sampling of the illumination sequence truncates the diffuse halo as displayed by the original RPSFs \rev{(Fig.~5g)}. This lack of information alters the aberration correction process and accounts for the residual background displayed by the final RPSFs \rev{(Fig.~5b)}. This academic experiment thus highlights both the benefits and limits of RMI. 



%

\end{document}